\documentclass[final,5p,times]{elsarticle}

\pdfoutput=1

\usepackage{graphicx}
\usepackage{amssymb}
\usepackage{amsmath}
\usepackage{bm}
\usepackage{lineno}
\usepackage{caption}
\usepackage{subcaption}
\usepackage[mathscr]{euscript}

\usepackage{amsthm}
\usepackage{color}
\usepackage{natbib}
\usepackage{titlesec}
\usepackage{xcolor}
\usepackage{float}
\usepackage{hyperref}

\begin{document}

%% -----------------------------------------------------------------------------------------------
%% -----------------------------------------------------------------------------------------------
\begin{frontmatter}

  \title{Charged particle track reconstruction with S$\pi$RIT Time Projection Chamber}

  %core group
  \author[korea]         {J.W. Lee\corref{cor}} \ead{ejungwoo@korea.ac.kr}
  \author[nscl]          {G.   Jhang\corref{cor}} \ead{changJ@nscl.msu.edu}
  \author[nscl]          {G.   Cerizza}
  %1st
  \author[nscl,msu]      {J.   Barney}
  \author[nscl,msu]      {J.   Estee}
  \author[riken]         {T.   Isobe}
  \author[kyu,riken]     {M.   Kaneko}
  \author[riken]         {M.   Kurata-Nishimura}
  \author[nscl,msu]      {W.G. Lynch}
  \author[kyu]           {T.   Murakami}
  \author[nscl,msu]      {C.Y. Tsang}
  \author[nscl,msu]      {M.B. Tsang}
  \author[nscl]          {R.   Wang}
  %2nd group
  \author[korea]         {B.   Hong}
  \author[tamu]          {A.B. McIntosh}
  \author[riken]         {H.   Sakurai}
  \author[nscl]          {C.   Santamaria}
  \author[nscl]          {R.   Shane}
  \author[nscl,msu]      {S.   Tangwancharoen}
  \author[tamu]          {S.J. Yennello}
  \author[tsu]       {and Y.   Zhang}
  \author{\\For the S$\pi$RIT Collaboration}

  \cortext[cor]{Corresponding authors}
  \address[korea]{Department of Physics, Korea University, Seoul 02841, Republic of Korea}
  \address[nscl]{National Superconducting Cyclotron Laboratory, Michigan State University, East Lansing, MI 48824, USA}
  \address[msu]{Department of Physics and Astronomy, Michigan State University, East Lansing, MI 48824, USA}
  \address[riken]{RIKEN Nishina Center, RIKEN, 2-1 Hirosawa, Wako, Saitama 356-0198, Japan}
  \address[kyu]{Department of Physics, Kyoto University, Kyoto 606-8502, Japan}
  \address[tamu]{Cyclotron Institute, Texas A\&M University, College Station, TX 77843, USA}
  \address[tsu]{Department of Physics, Tsinghua University, Haidian DS, Beijing 100084, China}

  \begin{abstract}
    In this paper, we present a software framework, S$\pi$RITROOT, which is capable of track reconstruction and analysis of heavy-ion collision events recorded with the S$\pi$RIT time projection chamber. The track-fitting toolkit GENFIT and the vertex reconstruction toolkit RAVE are applied to a box-type detector system. A pattern recognition algorithm which performs helix track finding and handles overlapping pulses is described. The performance of the software is investigated using experimental data obtained at the Radioactive Isotope Beam Facility (RIBF) at RIKEN. This work focuses on data from $^{132}$Sn + $^{124}$Sn collision events with beam energy of 270 AMeV. Particle identification is established using $\left<dE/dx\right>$ and magnetic rigidity, with pions, hydrogen isotopes, and helium isotopes.
  \end{abstract}


\end{frontmatter}
%% -----------------------------------------------------------------------------------------------
%% -----------------------------------------------------------------------------------------------

%%%%%%%%%%%%%%%%%%%%%%%%%%%%%%%%%%%%%%%%%%%%%%%%%%%%%%%%%%%%%%%%%%%%%%%%%%%%%%%%%%%%%%%%%%%%%%%%%%
%%%%%%%%%%%%%%%%%%%%%%%%%%%%%%%%%%%%%%%%%%%%%%%%%%%%%%%%%%%%%%%%%%%%%%%%%%%%%%%%%%%%%%%%%%%%%%%%%%
\section{Introduction}
The bulk properties of nuclear matter are described by the nuclear equation of state (EoS)~\citep{horowitz2014}.
Uncertainty in the nuclear EoS arises due to the symmetry energy contribution, which results from isospin asymmetry, defined as $\delta = (N-Z)/(N+Z)$, with $N$ and $Z$ being the number of neutrons and protons of the system. The SAMURAI Pion-Reconstruction and Ion-Tracker (S$\pi$RIT)
Time Projection Chamber (TPC)~\citep{shane2015} was designed to study the
symmetry energy of the compressed nuclear matter at the Radioactive Isotope Beam Factory (RIBF) at RIKEN. Promising observables for symmetry energy are collective flow~\citep{russotto2011}
of the charged particles and $\pi^-/\pi^+$ yield ratios~\citep{tsang2017}.

Data collected by TPCs are unique to the event types,
detector properties, and the DAQ system and electronics employed for the detector.
It is essential to develop a reconstruction software
fit for each unique experiment~\citep{ayyad2018,anderson2003,kouzinopoulos2016}.
S$\pi$RITROOT is a software framework developed to analyze for a series of S$\pi$RIT-TPC experiments.
It is based on the FairRoot~\citep{fairroot} framework and ROOT~\citep{root} package
and is composed of two main parts: one for simulation and the other for reconstruction.
The simulation part using Monte Carlo events with GEANT4 will be described in a future paper.
The main focus of this paper is on the reconstruction of tracks.

\begin{figure*}[ht]
  \centering\includegraphics[width=10cm]{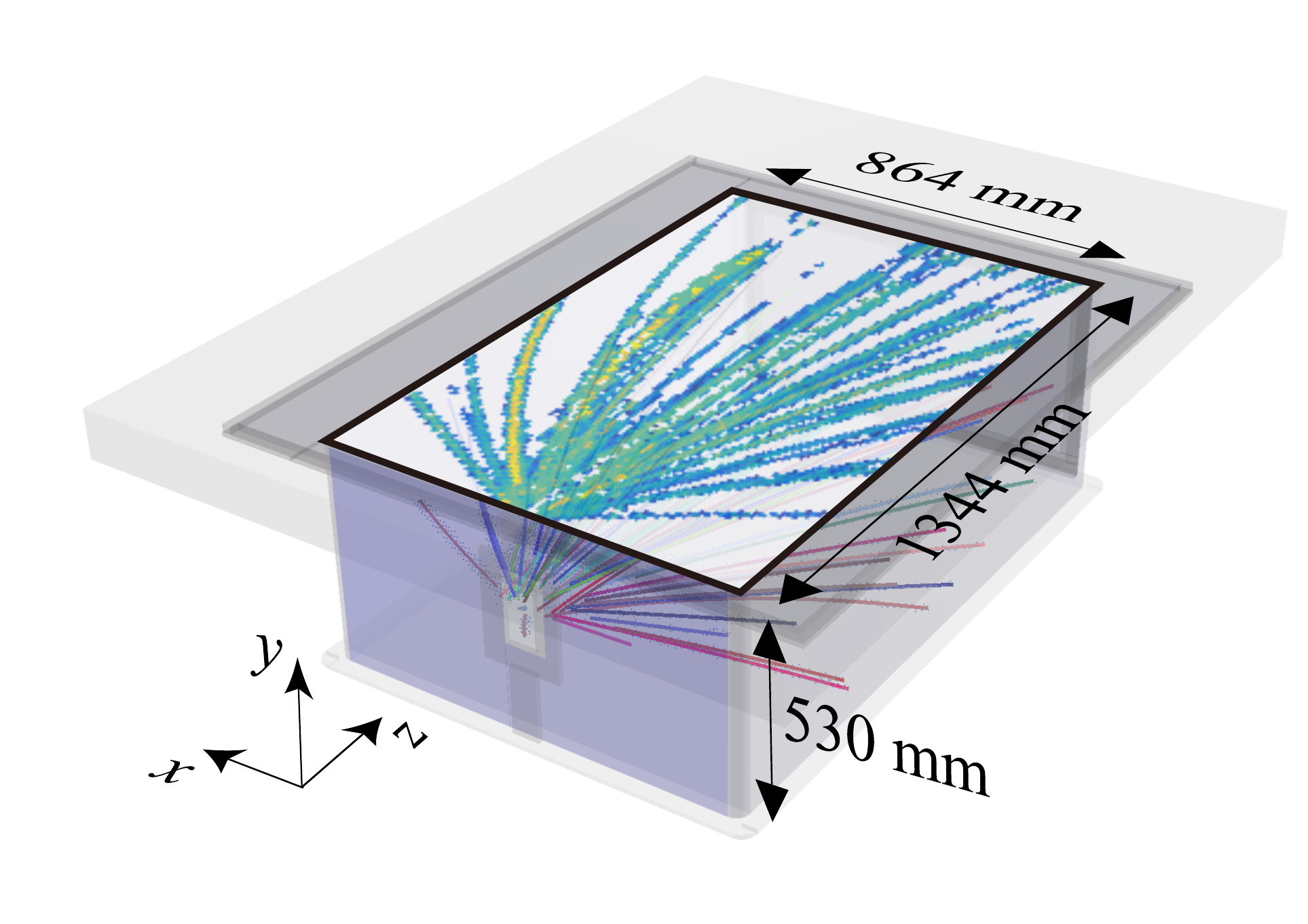}
  \caption{Event display of a typical $^{132}$Sn+$^{124}$Sn collision event.
  The pad plane on the top of the TPC is shown with the amount of charge of the event projected to each pad.
  The effective dimensions of the pad plane, and the height of the gas volume are indicated.
  The magnetic field is produced parallel to the negative $y$-axis
  while the electric field is anti-parallel to the magnetic field.}
  \label{figure:spirit-tpc}
\end{figure*}

Here we describe the process of reconstructing tracks from the raw data into information relevant to particle identification; specifically, the momentum vector, specific energy loss, and the collision vertex in events. We have adopted several open source
%{C\nolinebreak\hspace{-.05em}\raisebox{.4ex}{\tiny\bf +}\nolinebreak\hspace{-.10em}\raisebox{.4ex}{\tiny\bf +}\ }
C++
software packages into S$\pi$RITROOT such as GENFIT~\citep{rauch2015} for the momentum reconstruction and RAVE~\citep{waltenberger2011} for the vertex reconstruction. Although these packages have been mainly developed for cylindrical
detector systems~\citep{rauch2012}, we have successfully modified them to work with
the box-type detector system such as the S$\pi$RIT-TPC.

In April and May of 2016, Rare Isotope (RI) collision events were taken with S$\pi$RIT TPC inside
the SAMURAI magnet~\citep{sato2012}.
The main collision systems were
$^{132}$Sn+$^{124}$Sn,
$^{124}$Sn+$^{112}$Sn,\linebreak
$^{108}$Sn+$^{124}$Sn,
and $^{108}$Sn+$^{112}$Sn
at 270 AMeV with beam rate of 10 kHz.
The magnetic field produced by the SAMURAI magnet was 0.5 T.
The performance of the S$\pi$RITROOT software described
in this paper is based on the analysis of $^{132}$Sn+$^{124}$Sn data.

%%%%%%%%%%%%%%%%%%%%%%%%%%%%%%%%%%%%%%%%%%%%%%%%%%%%%%%%%%%%%%%%%%%%%%%%%%%%%%%%%%%%%%%%%%%%%%%%%%
%%%%%%%%%%%%%%%%%%%%%%%%%%%%%%%%%%%%%%%%%%%%%%%%%%%%%%%%%%%%%%%%%%%%%%%%%%%%%%%%%%%%%%%%%%%%%%%%%%
\section{S$\pi$RIT-TPC}
The S$\pi$RIT-TPC is a rectangular TPC operated with a Multi-Wire Proportional Chamber (MWPC) readout \cite{shane2015}.
The geometry and dimensions of the S$\pi$RIT TPC field cage is shown in Fig.~\ref{figure:spirit-tpc},
which also indicates the coordinate system used throughout this paper.
The wires run parallel to the $x$-axis, and are located just below the pad plane,
which is situated at the top of the field cage.
The pad plane consists of 108 rows (each row runs along the $z$-axis)
and 112 layers (each layer runs along the $x$-axis) of pads.
Each pad is 7.5 mm in $x$, and 11.5 mm in $z$, with a 0.5 mm gap between neighboring pads.
The target is centered at 13.2 mm upstream of the pad plane, 205 mm below the pad plane.

The Generic Electronics for TPC (GET) readout system is employed to read out signals from the pad plane\citep{isobe2018, pollacco2018}.
The charge on each pad is read out with a 25 MHz sampling rate,
corresponding to a 40 ns width for each sample, which we call a time bucket (tb).
Each channel can store up to 512 tb per event. The shaping time of the amplifier is set to 117~ns.
The dynamic range was chosen to be 120 fC so that the $\left<dE/dx\right>$
range covers pions, hydrogen isotopes, helium and lithium isotopes after software corrections~\citep{estee2019162509}.

The TPC field cage is filled with P10 gas, a mixture of Ar(90\%) and CH$_4$(10\%),
with a magnetic field of 0.5 T oriented along the positive $y$-axis and an electric field of 124.7 V/cm oriented along the negative $y$-axis.
The electron drift velocity in such conditions was calculated to be
5.54 cm/$\mu$s using MAGBOLTZ v11.9~\citep{biagi1999} simulations.
The experimental value for drift velocity was obtained by comparing
the measured $y$-position difference between TPC cathode plate and
beam with the corresponding time difference measured by the readout electronics,
resulting in a measured drift velocity around 5.54(1) cm/$\mu$s for the $^{132}$Sn + $^{124}$Sn system.

%%%%%%%%%%%%%%%%%%%%%%%%%%%%%%%%%%%%%%%%%%%%%%%%%%%%%%%%%%%%%%%%%%%%%%%%%%%%%%%%%%%%%%%%%%%%%%%%%%
%%%%%%%%%%%%%%%%%%%%%%%%%%%%%%%%%%%%%%%%%%%%%%%%%%%%%%%%%%%%%%%%%%%%%%%%%%%%%%%%%%%%%%%%%%%%%%%%%%
\section{Pulse Shape Analysis}

\begin{figure}[ht]
  \centering\includegraphics[width=6cm]{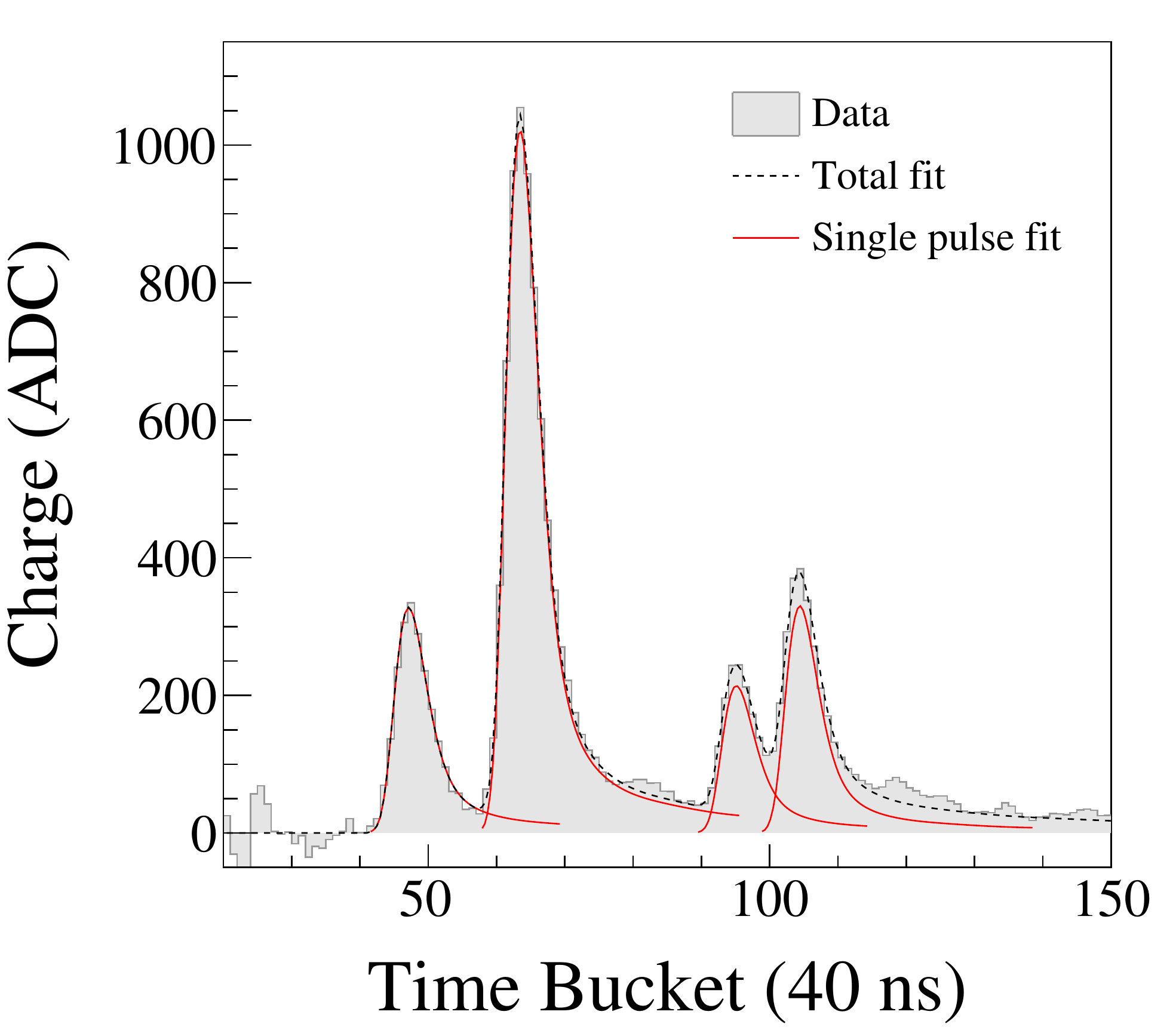}
  \caption{Result of the PSA with a pad containing multiple signals.
  The data in the gray histogram is fitted with multiple reference pulses shown as red functions.
  The sum of single-pulse fits is plotted as dashed black curve.}
  \label{figure:psa}
\end{figure}

The S$\pi$RIT-TPC measures events with charged particle track multiplicity ranging from 30 to 70 tracks, which form a common vertex near the target region.
The track density in the region near the target can be quite high; therefore,
pads in this region likely contains the sum of several overlapping
signals from several tracks, as shown in Fig.~\ref{figure:psa}.
The fast pulse fitting algorithm described in this section is developed to analyze such complex overlapping tracks.

The pulse shape was found to be stable regardless of particle type, drift length, and data type (i.e., cosmic data or nuclear collision data). Therefore a standard reference pulse shape was extracted from pads that contained only a single pulse with amplitude ranging from 1000 to 3000 ADC channels. The peak amplitudes of each pulse were normalized to one, and the peak position in time (of all the pulses) were aligned to an arbitrary reference time. After calculating the average of the normalized pulses, a reference pulse shape, $f_\textrm{ref}(\textrm{tb})$, was calculated with which we could use to fit the time bucket spectra of each pad.

The start time of a hit in the TPC is defined to be the time at which the pulse achieves 5 \% of its peak value.
The measured peaking time of the reference pulse is around 4.3 tb (172 ns) which defines the time duration from 5 \% to 100 \% of its peak amplitude.

In the fast pulse fitting algorithm, a pulse is parameterized by its amplitude and start time.
The algorithm works in a time sequence; starting from the first time bucket and stepping through the time spectrum
until a pulse is found by identifying a peak.
The spectrum is then fitted using the reference pulse in the peaking time range
with the minimum $\chi^2$ fit method.
The fitted pulse is subtracted from the raw spectrum.
This process is repeated until the end of the time buckets.

The single hit finding efficiency, given below in Eq.~\ref{eq:1hit}, is calculated by comparing the
total number of pads that a track pass through directly, $n_\textrm{total}$,
and the number of hits reconstructed in a track among these pads, $n_\textrm{reco}$.
The efficiency is found to be $\epsilon_\textrm{1-hit} = 95\pm1\,\%$.
\begin{equation}
  \epsilon_\textrm{1-hit} =
 \frac{n_\textrm{reco}}{n_\textrm{total}}.
 \label{eq:1hit}
\end{equation}

\begin{figure}[ht]
  \centering\includegraphics[width=6cm]{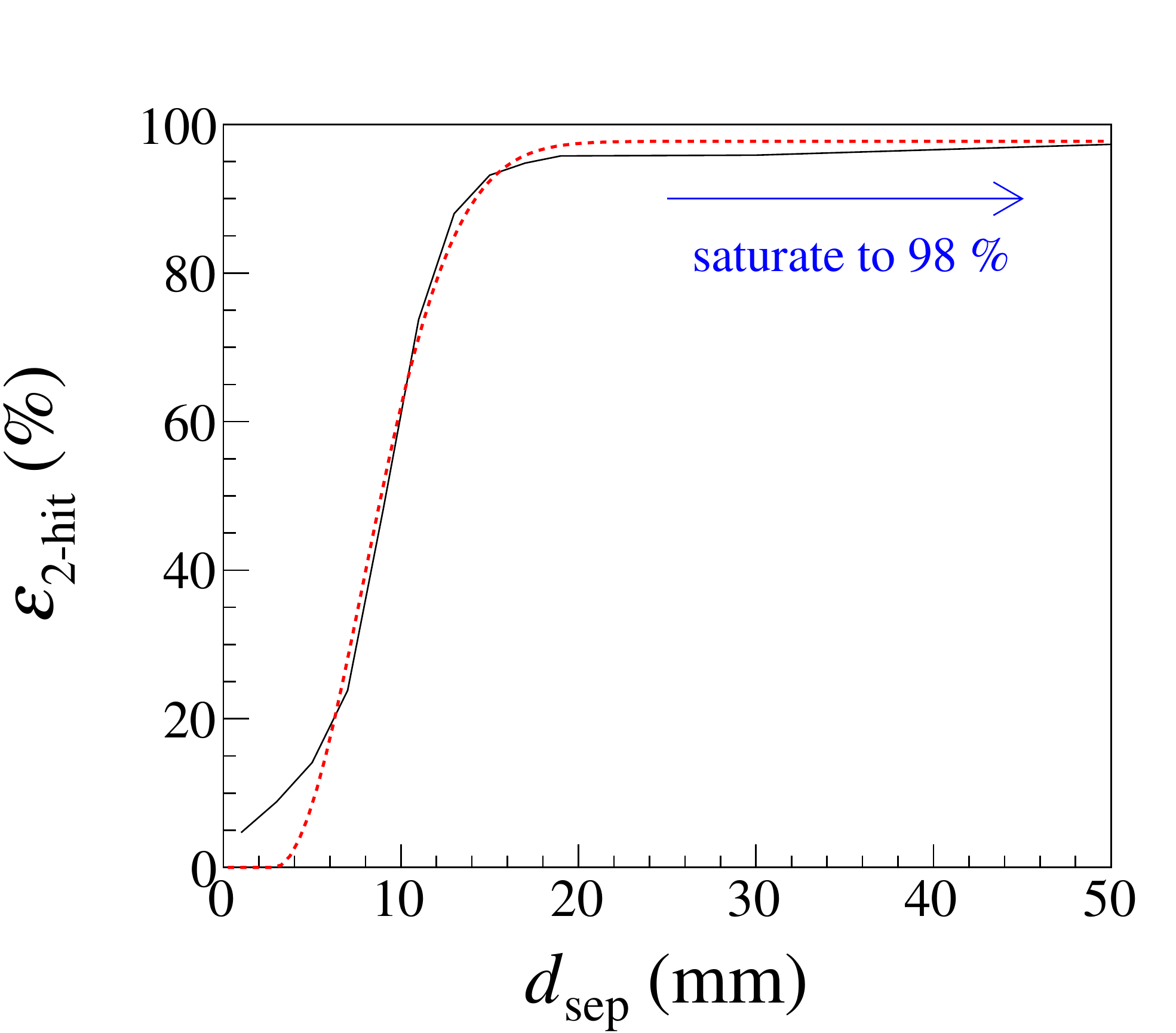}
  \caption{Two hit separation efficiency $\epsilon_\textrm{2-hit}$
  plotted as a function of the separation distance $d_\textrm{sep}$.
  The data is fitted with Eq.~\ref{eq:2hit_eff_fit} (dotted red) to guide the eye.
  The region for $d_\textrm{sep} \leq 20$ mm shows a large variance in efficiency
  while region for $d_\textrm{sep} > 20$ mm is saturated.
  }
  \label{figure:hit_separation}
\end{figure}

By checking for reconstructed hits of two track events, we can count the number of pads which reconstruct both hits, $n_{1,2}$, to the number of pads containing only the first hit, $n_{1}$. The two-hit separation efficiency is then defined as
\begin{equation}
  \epsilon_\textrm{2-hit} =
  \frac{1}{\epsilon_\textrm{1-hit}}\times\frac{n_{1,2}}{n_{1}}.
  \label{eq:2hit}
\end{equation}
The hit separation distance is calculated using information from pads that should contain pulses from two reconstructed tracks. The expected arrival time at the pad plane of the first and second pulses are calculated from the track fit parameters and referred to as $y_1$ and $y_2$, where $y_1 > y_2$. The difference between these values defines the hit separation distance, $d_\textrm{sep} = y_1 - y_2$. The two-hit separation efficiency given by Eq.~\ref{eq:2hit} is estimated as a function of the hit separation distance between two consecutive pulses, $d_\textrm{sep}$, and is shown in Fig.~\ref{figure:hit_separation}.
The data in Fig.~\ref{figure:hit_separation} is fitted with
\begin{equation}
  \epsilon_\textrm{2-hit} = \epsilon_\textrm{max} \left(1\!-\!\exp\frac{-(d_\textrm{sep}\!-\!d_{0})^{2}}{\delta_{d}^{2}}\right),
  \label{eq:2hit_eff_fit}
\end{equation}
where $\epsilon_\textrm{max}$, $d_{0}$ and $\delta_{d}$ are the fit parameters.
The efficiency saturates to $\epsilon_\textrm{max} =  98\,\%$ because of the uncertainty in the reference pulse shape.
Due to low efficiencies to identify multiple tracks, we do not reconstruct tracks in the region with the highest track densities near the target.

%%%%%%%%%%%%%%%%%%%%%%%%%%%%%%%%%%%%%%%%%%%%%%%%%%%%%%%%%%%%%%%%%%%%%%%%%%%%%%%%%%%%%%%%%%%%%%%%%%
%%%%%%%%%%%%%%%%%%%%%%%%%%%%%%%%%%%%%%%%%%%%%%%%%%%%%%%%%%%%%%%%%%%%%%%%%%%%%%%%%%%%%%%%%%%%%%%%%%
\section{Track Finding}
Finding tracks from a set of hits collected by the PSA task requires pattern recognition (PR) algorithms to sort and group all the hits that belong to a particular track.
A charged particle moving inside a magnetic field follows a helical trajectory; thus we limit our search to finding helical tracks inside the TPC.
Finding which subset of hits makes a unique helix is a massive comparison process,
therefore a fast helix tracking method was developed
by implementing a Riemann circular fit method~\citep{frhwirth2018,strandlie2000},
and a Point Of Closest Approach (POCA) approximation.

We introduce the helix track parameterization and POCA approximation in section~\ref{helix_par},
the overall work flow of track building in section~\ref{track_finding_algorithm},
and the fast helix fit method in section~\ref{helix_fit}.

%%%%%%%%%%%%%%%%%%%%%%%%%%%%%%%%%%%%%%%%%%%%%%%%%%%%%%%%%%%%%%%%%%%%%%%%%%%%%%%%%%%%%%%%%%%%%%%%%%

\subsection{Helix track parameterization} \label{helix_par}
We parameterize a closed-end helix track with the central-axis in the $y$ direction by the following 9 parameters:
\begin{itemize}
  \item Helix center $\boldsymbol x_\textrm{h}=(x_\textrm{h},z_\textrm{h})$ of the circle projected onto the pad plane,
  \item Helix radius $R_\textrm{h}$ of the projected circle,
  \item Polar angle of the helix $\alpha$ being the parametric angle defining the $x$ and $z$ values (values at the two end points of the helix are defined as $\alpha_\textrm{tail}$ and $\alpha_\textrm{head}$, respectively):
  \begin{eqnarray}
  &&x=R_\textrm{h} \cos\alpha + x_\textrm{h},\\
  &&z=R_\textrm{h} \sin\alpha + z_\textrm{h},
  \end{eqnarray}
  \item Slope parameter $s_\textrm{h}$ and $y$-offset $y_\textrm{h}$, which describe
  the linear relation between the polar angle $\alpha$ and the $y$-position of the helix:
  \begin{equation}
    \label{eq:alpha_y_relation}
    y = s_\textrm{h} \alpha + y_\textrm{h},
  \end{equation}
  \item The track window $\Delta_{\rho}$ and $\Delta_{\tau}$ of the two axis $\rho$ and $\tau$,
  which are described below.
\end{itemize}

The dip angle describing the angle between the track momentum and the pad plane is defined by
\begin{equation}
  \theta=\arctan{\left(\frac{s_\textrm{h}}{R_\textrm{h}}\right)}.
\end{equation}

\begin{figure}[ht]
  \centering\includegraphics[width=5cm]{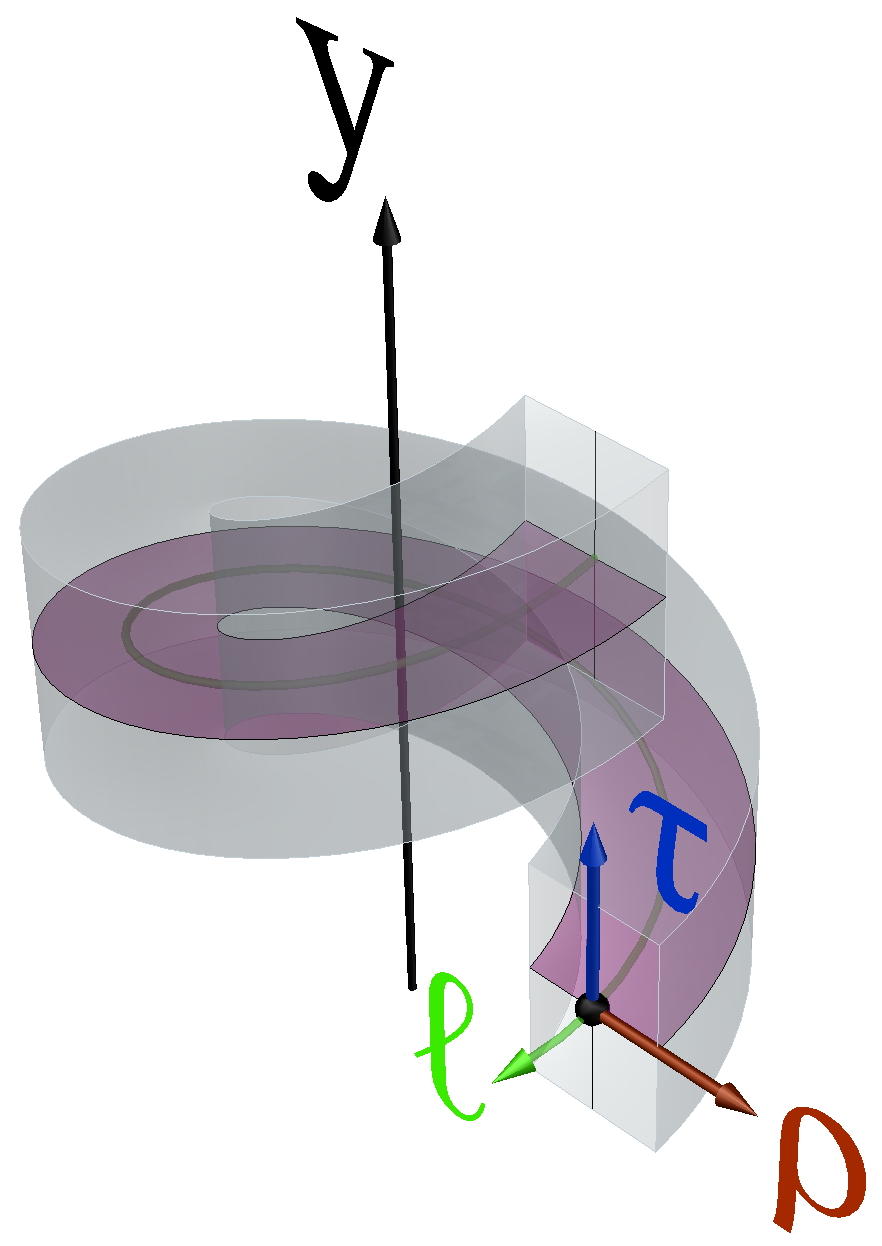}
  \caption{Definition of the helicoid coordinate $(\rho,\tau,\ell)$.
  The helix track in green defines the $\hat{\boldsymbol\ell}$-axis,
  $\hat{\boldsymbol\rho}$ directs the radial axis,
  and $\hat{\boldsymbol\tau}$ is normal to the helicoid plane shown in purple.}
  \label{figure:helicoid_coordinate}
\end{figure}

To simplify helix track parameterization,
we employ the helicoid coordinate system as shown in Fig. \ref{figure:helicoid_coordinate}.
The helicoid coordinate is defined by choosing
the first axis to be the radial axis $\hat{\boldsymbol\rho}$ from the helix center,
the second axis as the helix path $\hat{\boldsymbol\ell}$,
and the last axis $\hat{\boldsymbol\tau}$ is normal to both
$\hat{\boldsymbol\ell}$ and $\hat{\boldsymbol\rho}$.
The transformation from the point $\boldsymbol{x_\textrm{i}}$ with corresponding $\alpha_\textrm{i}$ is given by
\begin{eqnarray}
  \label{eq:helix_map}
  \rho_\textrm{i} &=& R_\textrm{i} \!- R_\textrm{h},\\
  \tau_\textrm{i} &=& \delta y_\textrm{i}\sec{\theta},\\
  \ell_\textrm{i} &=& \alpha_\textrm{i} R_\textrm{h}\sec{\theta}
  \!+ \delta y_\textrm{i}\sin{\theta},
\end{eqnarray}
where $R_\textrm{i} = \sqrt{\left(x_\textrm{i}\!-x_\textrm{h}\right)^2 \!+ \left(z_\textrm{i}\!-z_\textrm{h}\right)^2}$
and $\delta y_\textrm{i} = y_\textrm{i}\!-s_\textrm{h}\alpha_\textrm{i}\!+ y_\textrm{h}$.

Here we approximate POCA to $(\rho,\tau,\ell)_\textrm{poca}=(0,0,\ell_\textrm{i})$ so that the values $\rho_\textrm{i}$ and $\tau_\textrm{i}$ correspond to the shortest distance from the track in each axis.
One can calculate the closest distance directly from these values,
but it is more useful to express them by separate parameters.
The Root Mean Square (RMS) values $\textrm{RMS}(\rho)_\textrm{track}$ and $\textrm{RMS}(\tau)_\textrm{track}$
of hits in a track refer directly to the window of the helix track as given in Eqs.~\ref{eq:track_window_rho} and \ref{eq:track_window_tau}.
These RMS values typically range from 1 to 5 mm.
\begin{eqnarray}
  \label{eq:track_window_rho}
  \Delta_\rho &=& 3.5\times\textrm{RMS}(\rho)_\textrm{track},\\
  \label{eq:track_window_tau}
  \Delta_\tau &=& 3.5\times\textrm{RMS}(\tau)_\textrm{track}.
\end{eqnarray}

%%%%%%%%%%%%%%%%%%%%%%%%%%%%%%%%%%%%%%%%%%%%%%%%%%%%%%%%%%%%%%%%%%%%%%%%%%%%%%%%%%%%%%%%%%%%%%%%%%

\subsection{Track building algorithm} \label{track_finding_algorithm}

Typically the collision events in the S$\pi$RIT-TPC have a high track density around the target position, whereas at the far side of the TPC, the track density is much lower due to the outgoing emission angles and the presence of the magnetic field
that separates the tracks by their magnetic rigidity. The challenging task of event reconstruction can be simplified by starting the tracking at the end of the tracks and iteratively working back toward the collision vertex.

\begin{figure*}[ht]
  \centering\includegraphics[width=11cm]{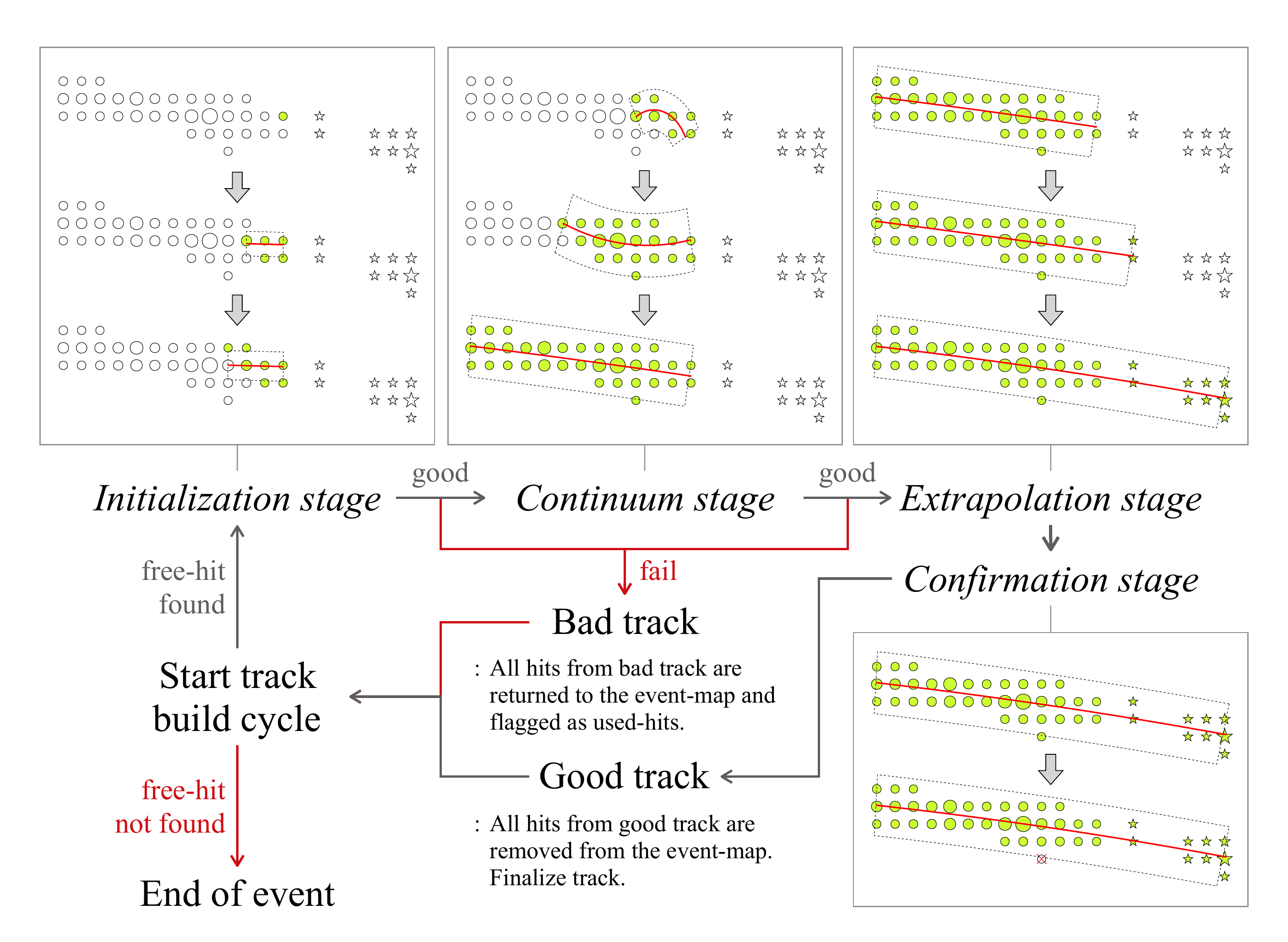}
  \caption{Flow chart of the track building algorithm with each stage showing an example track evolution.
  The example cartoons demonstrate the case of a successfully built track.
  In the figures, free-hits are shown as circles and used-hits are shown as stars.
  The used-hits originate from the earlier track buildings that have failed.
  The track-hits, filled in green, belong to the track being built, and are fit with a straight line for the initialization stage,
  then fit with helical parameterization for the other stages. The area surrounded by the dotted line shows the track window $\Delta_{\rho}$.
  See section~\ref{track_finding_algorithm} for further details.
  }
  \label{figure:track_finding_flow}
\end{figure*}

The flow chart of the track building algorithm is shown in Fig.~\ref{figure:track_finding_flow}.
For the first step, the event-map containing all reconstructed hits is created.
The event-map is sophisticated in that for a given position it will find the corresponding pad and its
neighboring pads, where each of those pads contains reconstructed hits.

The hits are classified into three group of flags:
free-hits, which have not participated in track building,
used-hits, which participate in track building,
and track-hits, which have been collected as part of a track.
While initially all hits are flagged as a free-hit, hits that are collected during the process of track building are flagged as a track-hit.
After a track is successfully built, track-hits are removed from the event-map.
If the track-build fails, track-hits are returned to the event-map and flagged as used-hit.
This type of flag is essential to prevent track building from starting in island of isolated hits
which cannot be built into a good track:
e.g., the used-hits of star markers in Fig.~\ref{figure:track_finding_flow}.

In this track building algorithm only one track is built at a time.
The first hit of a track is selected by choosing the farthest hit from the target among free-hits in the event-map.
The track is then built in four stages:
\textit{Initialization}, \textit{Continuum}, \textit{Extrapolation} and \textit{Confirmation}.
In each stage, candidate hits are searched from the neighboring pads and from pads that are near the extrapolated track.
If a candidate hit falls into the track window $\Delta_{\rho,\tau}$,
the hit is added to the track and the track parameters are updated by track fitting.

In the \textit{Initialization} stage, hits from neighboring pads are collected
and the track length is calculated with linear fit of the track.
This stage fails if the track length is smaller than $2.5\times\textrm{RMS}(\rho)_\textrm{track}$ with at most 15 hits.

In the \textit{Continuum} stage, all hits from neighboring pads that fit into the track window $\Delta_{\rho,\tau}$ are added.
The area surrounded by the dotted line in Fig.~\ref{figure:track_finding_flow}
shows the range of track window $\Delta_{\rho}$.
The radius of the helix must be larger than 25 mm at the end of this stage.

The \textit{Extrapolation} stage works in the same way as the \textit{Continuum} stage,
but the candidate hits are added by extrapolating the track across regions where there may be no hits, until the track extrapolation reaches the boundary of TPC volume. Discontinuities in the track can be caused by low gain regions or saturated regions where little to no charge was extracted, resulting in a broken track.

Finally, in the \textit{Confirmation} stage, the hits are compared with the helix parameter set
until no more hits are removed or added.

After a track is built, a new cycle begins with remaining free-hits in the event-map.
The process is repeated until the event-map no longer contains free-hits.

%%%%%%%%%%%%%%%%%%%%%%%%%%%%%%%%%%%%%%%%%%%%%%%%%%%%%%%%%%%%%%%%%%%%%%%%%%%%%%%%%%%%%%%%%%%%%%%%%%

\subsection{Fast helix fit via Riemann fit} \label{helix_fit}
The helix parameters introduced in section~\ref{helix_par} are used in many places throughout the software.
For example, it will be shown in section~\ref{hit_clusterization}
that the guide line of hit clusterization is based on the path and direction of the track,
which are determined with the helix parameters.
Also, the parameters are used as the initial values of momentum tracks are fitted with GENFIT.
Most importantly, an update of the track window is required in every track building step
which is why the fast helix fit is crucial in this software.

In this section we describe fast helix fit via Riemann circle fit which can find the
three circle parameters $\boldsymbol x_\textrm{h}=(x_\textrm{h},z_\textrm{h})$ and $R_\textrm{h}$ among the helix parameter set.
The slope and offset parameters $s_\textrm{h}$ and $y_\textrm{h}$ can be determined with a standard linear fit
and the remaining parameters, the polar angles at the end of helix ($\alpha_{head}$ and $\alpha_{tail}$), can be found with the POCA method.

\begin{figure}[ht]
  \centering\includegraphics[width=6cm]{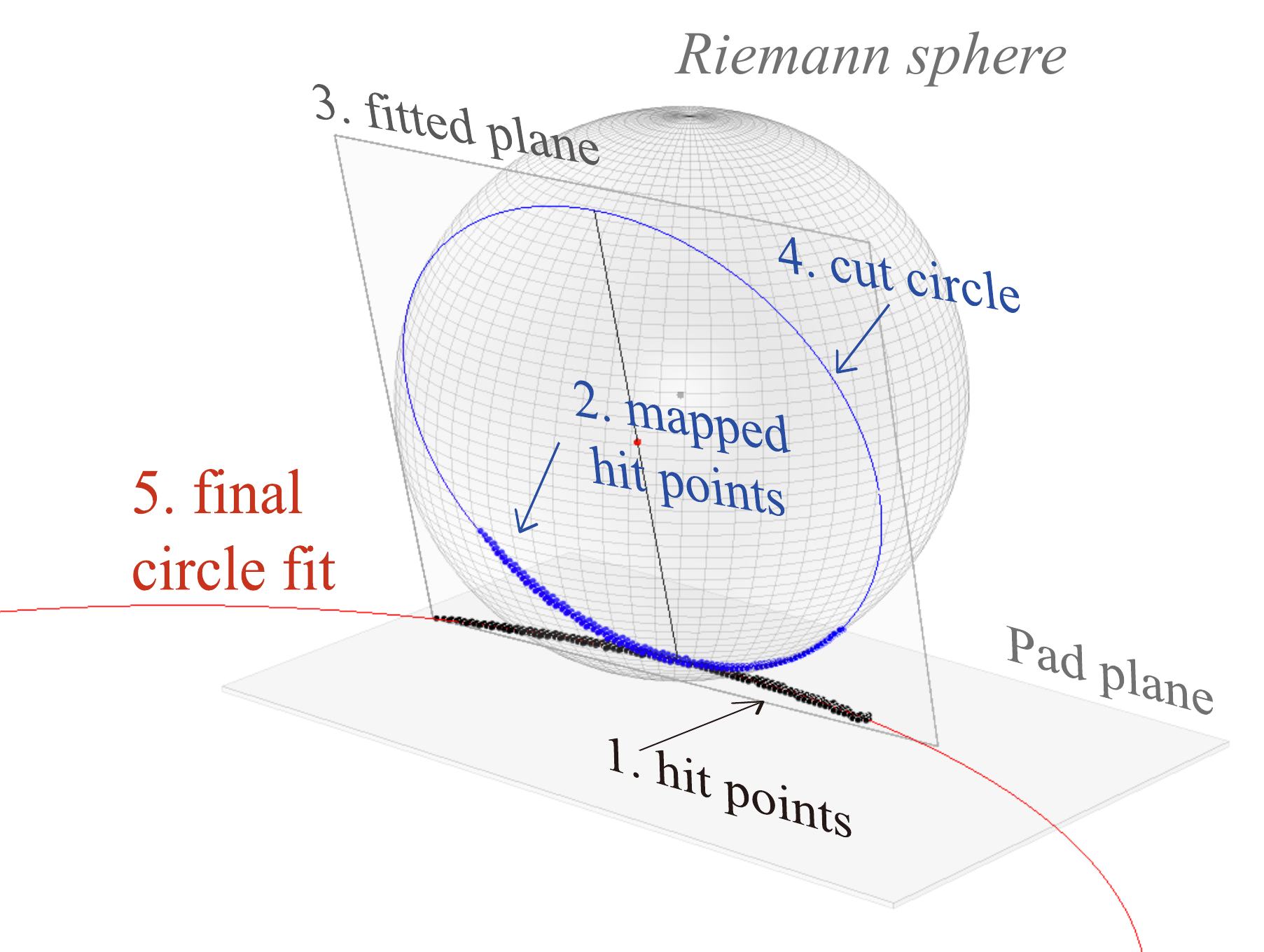}
  \caption{The steps of the Riemann fit process are shown with the Riemann sphere sitting on the pad plane.
  Firstly, the black hit points (1) on the pad plane are
  projected onto the Riemann sphere surface as shown by the blue points (2).
  Next, a plane is fitted to the projected points (3).
  The intersection of the Riemann sphere and fit-plane defines the blue circle (4).
  Finally, the blue circle is projected back to the pad plane, resulting in the red circle line (5).}
  \label{figure:Riemann_sphere}
\end{figure}

The $\chi^2$ fit of the circle is known as a non-linear problem.
We simplify the problem by implementing the so-called Riemann circle fit~\citep{frhwirth2018,strandlie2000}.
This method uses the fact that the plane fit is a linear problem
and that the circle in the pad plane uniquely maps onto a circle on the sphere.
The circle on a sphere uniquely defines a plane which can be solved by Orthogonal Distance Regression (ODR) method.
The sphere used in the Riemann circle fit which sits right on top of pad plane is called the Riemann sphere.
Figure \ref{figure:Riemann_sphere} shows the the explicit steps of the Riemann circle fit.

The center and radius of the Riemann sphere are chosen to be
$\boldsymbol{x}_\textrm{R} = \left<\boldsymbol x\right>_\textrm{hit} + R_\textrm{R}\,\hat{\boldsymbol{y}}$
and $R_\textrm{R} = 2 \boldsymbol{\sigma}_\textrm{hit}^2$, respectively,
where $\left<\boldsymbol x\right>_\textrm{hit}$ and $\boldsymbol{\sigma}_\textrm{hit}^2$
are the mean and variance of the hit position in the pad plane.
These parameters are chosen so that the projected points are widely distributed on the surface of the sphere.

The fit starts by mapping data points onto the Riemann sphere by stereo-graphic projection.
The intersection of the sphere and line connecting the hit to the north-pole of the sphere
defines the projected points $\boldsymbol{x}_\textrm{map}$,
which can be written as
\begin{eqnarray}
  \boldsymbol x_\textrm{map} = \frac{
  \left(\left[ x_\textrm{hit}\right],\, \frac{r_\textrm{eff}^2}{2R_\textrm{R}},\, \left[ z_\textrm{hit}\right] \right)
  }
  {{1+r_\textrm{eff}^2}}
  + \left<\boldsymbol x_\textrm{hit}\right>,
\end{eqnarray}
with mean-subtracted position
$\left[ \boldsymbol x \right]\!=\boldsymbol x\!-\left<\boldsymbol x\right>$
and effective radius
$r_\textrm{eff}=\left.\sqrt{\left[ x_\textrm{hit}\right]^2+\left[ z_\textrm{hit}\right]^2}\,\big/ 2R_\textrm{R}\right.$.

As the next step, a plane is fitted to the points on the Riemann sphere.
The solution is given by solving the eigenvalues and eigenvectors of the matrix
\begin{equation}
  \boldsymbol{A} = \sum_{i}^{N_\textrm{hit}} \boldsymbol X_i^\textrm{T}\boldsymbol X_i,
\end{equation}
where $\boldsymbol X_i = \sqrt{q_i} \left(\boldsymbol{x}_{\textrm{map},i}\!- \left<\boldsymbol{x}_\textrm{map}\right>\right)$ with hit charge $q$.
By choosing the smallest eigenvalue, the corresponding eigenvector
becomes the normal vector $\hat{\boldsymbol{n}}$ of the fit-plane $\mathscr{P}'$
defined by ${\hat{\boldsymbol{n}}\cdot\boldsymbol{x} = k}$,
where $k$ is the distance from the origin to the plane $\mathscr{P}'$.
The cross section of the the Riemann sphere and plane $\mathscr{P}'$ defines a cut circle $\mathscr{C}'$.

\begin{figure}[ht]
  \centering\includegraphics[width=6cm]{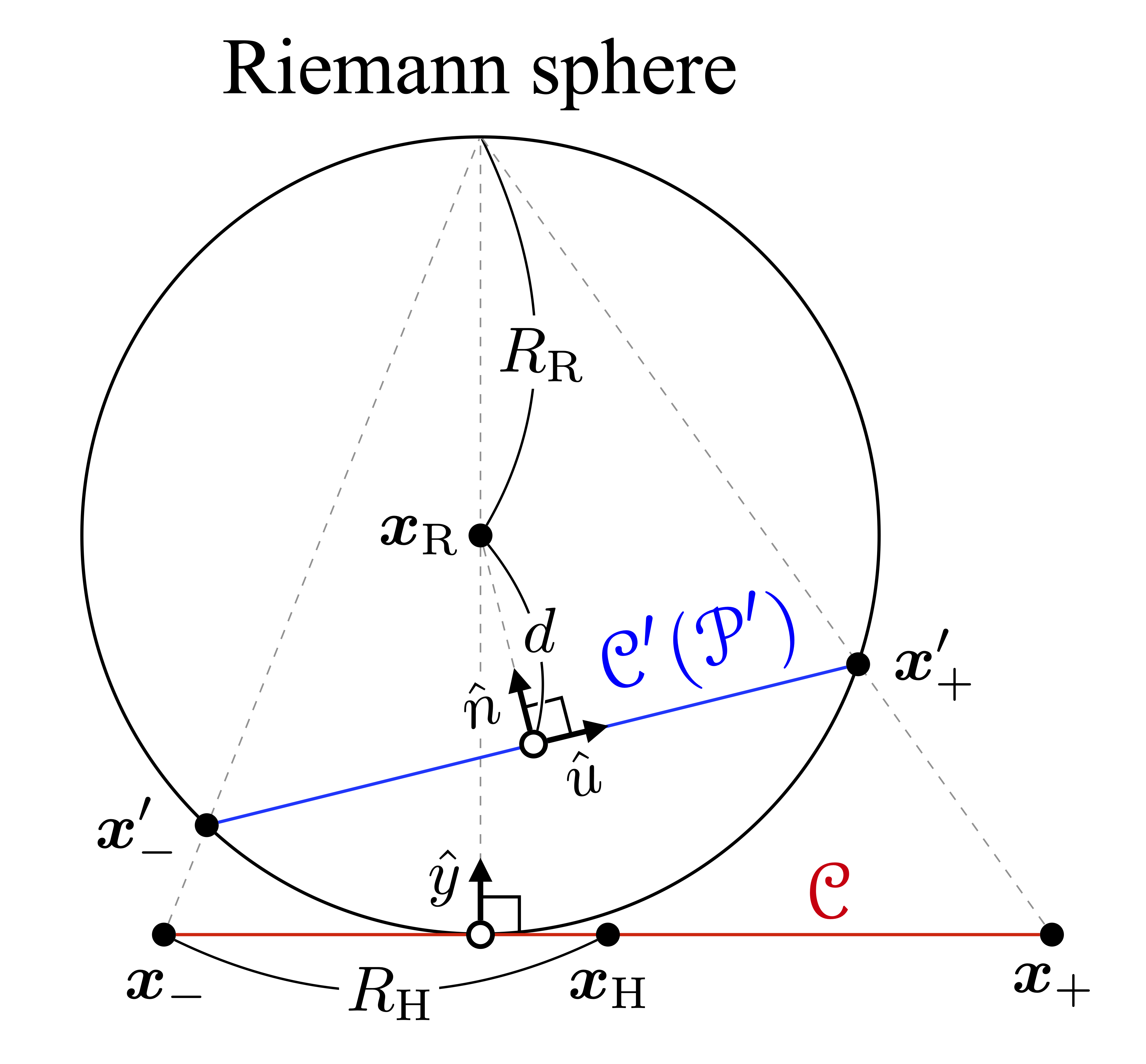}
  \caption{A cross-section of the Riemann sphere and the components used to evaluate
  the helix parameters in the plane of $\hat{\boldsymbol{n}}$ and $\hat{\boldsymbol{u}}$.
  The circle fit of the original hit distribution $\mathscr{C}$ is shown in red.
  The circle $\mathscr{C}'$ defined by the intersection of the Riemann sphere
  and plane $\mathscr{P}'$ is shown in blue.
  The inverse projection of points $\boldsymbol{x}_{\pm}'$ to $\boldsymbol{x}_{\pm}$
  are indicated.
  }
  \label{figure:riemann_parameters}
\end{figure}

The circle fit of the original hit distribution $\mathscr{C}$
is found by the inverse projection of a circle $\mathscr{C}'$.
The sketch of the inverse projection is shown in Fig.~\ref{figure:riemann_parameters}
to explain the geometrical meaning of each of the components in the paragraph.
We start by finding the highest-$y$ point $\boldsymbol{x}_{+}'$ and
lowest-$y$ point $\boldsymbol{x}_{-}'$ of a circle $\mathscr{C}'$.
The inverse projection of these two points $\boldsymbol{x}_{\pm}$
define the endpoints of the line that becomes the diameter of the circle $\mathscr{C}$.
Since the two eigenvectors $\hat{\boldsymbol{l}}$ and $\hat{\boldsymbol{m}}$
from the first and second largest eigenvalues of matrix $\boldsymbol{A}$ always lie in the plane $\mathscr{P}'$,
a unit vector $\hat{\boldsymbol{u}}$
pointing from the center of the circle $\mathscr{C}'$ to $\boldsymbol{x}_{+}$ is found by
\begin{eqnarray}
  \hat{\boldsymbol{u}} =
  \frac{1}{\sqrt{l_y^2+m_y^2}}\left(l_y\,\hat{\boldsymbol{l}}+m_y\,\hat{\boldsymbol{m}}\right).
\end{eqnarray}
Therefore, the points $\boldsymbol{x}_{\pm}'$ and $\boldsymbol{x}_{\pm}$ are written as
\begin{eqnarray}
  \boldsymbol{x}_{\pm}' &=& \boldsymbol{x}_\textrm{R} - d\hat{\boldsymbol{n}}
  \pm \sqrt{R_\textrm{R}^2 - d^2}\,\hat{\boldsymbol{u}},\\
  \boldsymbol{x}_{\pm} &=& \frac{1}{1-y_{\pm}'/2R_\textrm{R}}\left(x_{\pm}',0,z_{\pm}'\right),
\end{eqnarray}
where $d$ is the distance from the center of the Riemann sphere to the plane $\mathscr{P}'$;
$d = \left|\hat{\boldsymbol{n}}\cdot\boldsymbol{x}_\textrm{R}-k\right|$.
Finally, the helix parameters defined as
\begin{eqnarray}
  R_\textrm{h} &=& \left|\boldsymbol{x}_{+} - \boldsymbol{x}_{-}\right|/2,\\
  \boldsymbol{x}_\textrm{h} &=& \left(\boldsymbol{x}_{+} + \boldsymbol{x}_{-}\right)/2.
\end{eqnarray}

%%%%%%%%%%%%%%%%%%%%%%%%%%%%%%%%%%%%%%%%%%%%%%%%%%%%%%%%%%%%%%%%%%%%%%%%%%%%%%%%%%%%%%%%%%%%%%%%%%

\section{Hit Clusterization} \label{hit_clusterization}

\begin{figure}[ht]
\centering\includegraphics[width=6cm]{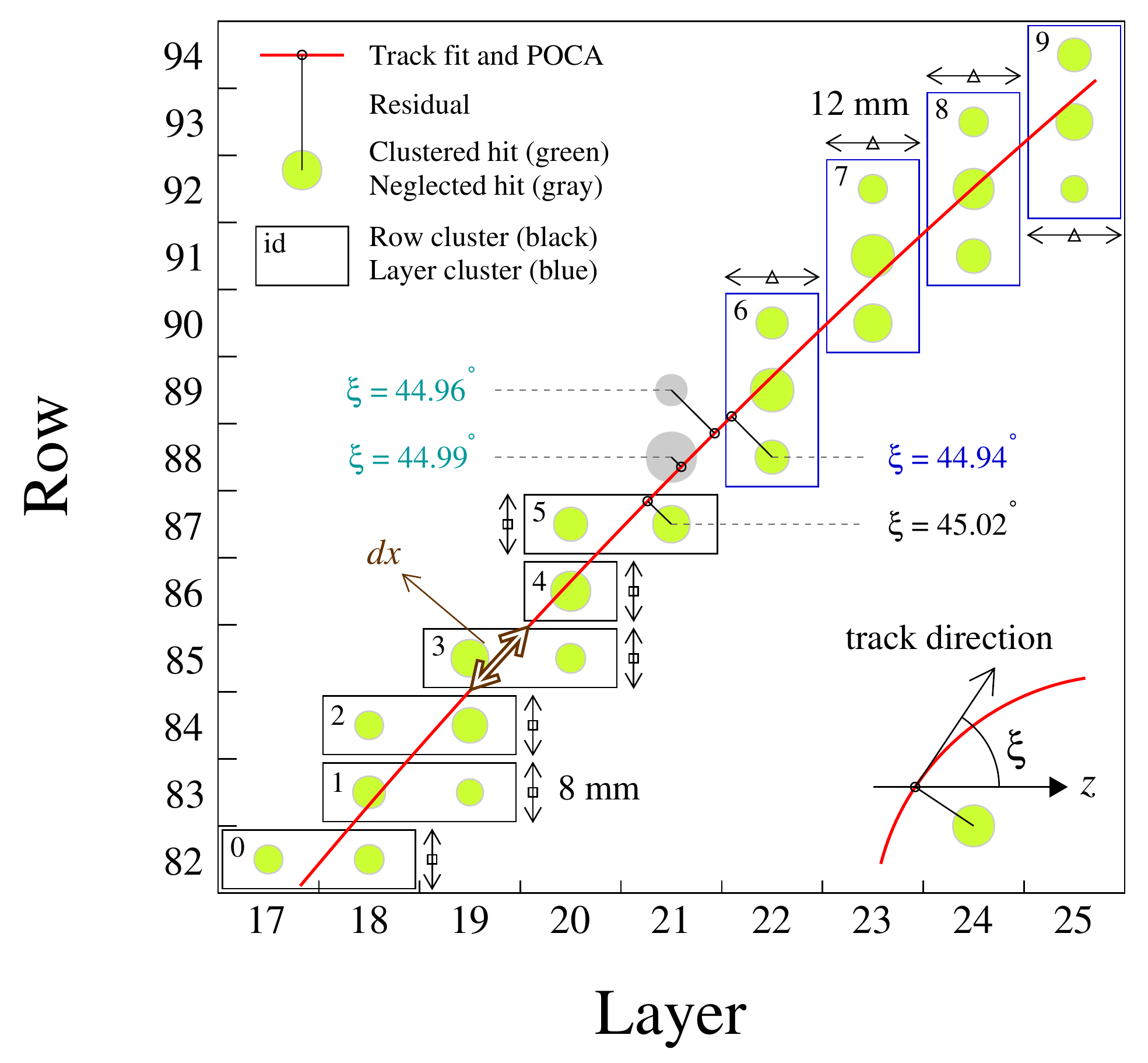}
\caption{An example track creating clusters of two different types.
The corresponding crossing angle $\xi$ of the hits is defined as cartoon on the bottom right corner.
Row-clusters are created from hits with $\xi > 45^\circ$
and layer-clusters are created from hits with $\xi < 45^\circ$.
Since the last row-cluster (5) and the first layer-cluster (6) should not share the same row or layer of pad,
two hits in between are neglected from the clusterization.
}
\label{figure:clusterization}
\end{figure}

The mean value of the hit distribution perpendicular to the track's trajectory gives the best estimate of the center of the track.
Although the pad size is only 7.5 mm $\times$ 11.5 mm,
we can significantly improve the position resolution, which is needed to reconstruct accurate momentum,
by grouping clusters of hits together,
either in the same row (row-cluster) or in the same layer (layer-cluster).

The classification of clusters are determined by the crossing angle of the track $\xi$,
defined as the angle between track momentum and the layer (or $z$-axis) of the pad plane
as shown in Fig.~\ref{figure:clusterization}:
\begin{eqnarray}
  \label{eq:cluster_type}
  \textbf{row-cluster}   :& 45^\circ < \left|\xi\right| < 135^\circ,\\
  \textbf{layer-cluster} :& \left|\xi\right| < 45^\circ, \,\,\,\,\left|\xi\right| > 135^\circ.
\end{eqnarray}
The hit cluster charge $Q$ is given by the sum of hit charges.
The hit cluster position (along the direction of clustering) is determined by averaging
charge-weighted positions of each hit in the cluster;
the perpendicular coordinate is set to the center of the pad.
For example, if hits are being clustered in a layer ($x$-direction),
the $x$-position is the weighted mean value of the hits
and the $z$-position is set to the $z$-position of the center of that layer, and vice versa for row clustering.

\begin{figure}[ht]
\centering\includegraphics[width=6cm]{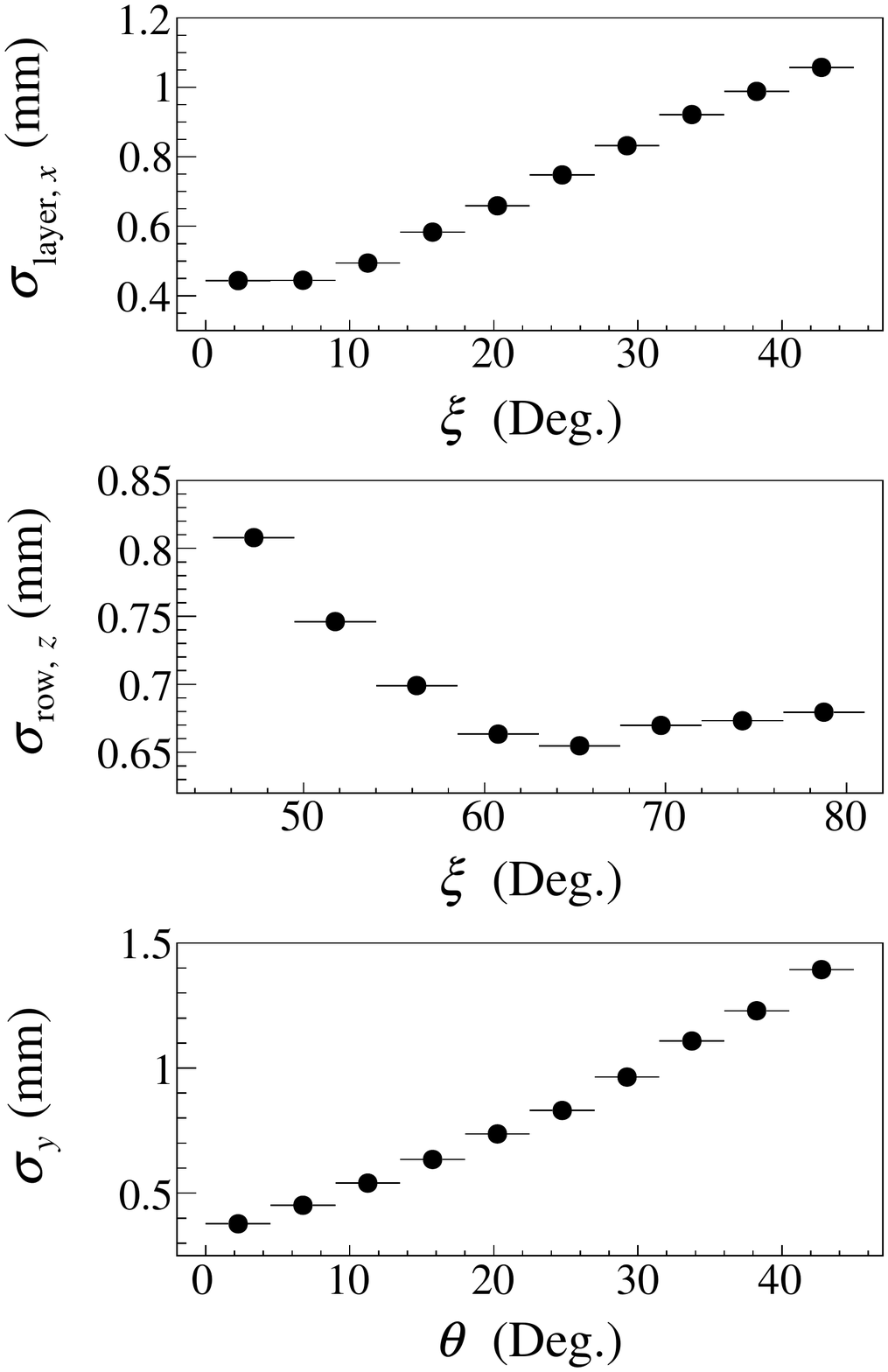}
\caption{Position errors of a hit cluster as functions of the
dip angle $\theta$ and the crossing angle $\xi$.
It is shown that $\sigma_{y}$ constantly increases as $\theta$ gets larger.
In addition, $\sigma_{\textrm{layer},x}$ and $\sigma_{\textrm{row},z}$ become larger
as $\xi$ gets close to $45^\circ$.
Saturation of $\sigma_{\textrm{row},z}$ at $\xi>65^\circ$
and $\sigma_{\textrm{layer},x}$ at $\xi<10^\circ$
can be seen which is an effect of the pad response function.
}
\label{figure:hit_cluster_position_error}
\end{figure}

The position error of each cluster was estimated from the residual
distribution of the cluster's position to the position of the reconstructed track.
The errors can be written separately depending on the type of hit clusters as given in Eqs.~\ref{eq:sigma} and \ref{eq:sigma_layer}.
\begin{eqnarray}
  \label{eq:sigma}
  \boldsymbol{\sigma}_\textrm{row}   &=& \left(1.73\,\, \textrm{mm},
  \,\,\sigma_{y}(\theta), \,\,\sigma_{\textrm{row},z}(\xi)\right),\\
  \label{eq:sigma_layer}
  \boldsymbol{\sigma}_\textrm{layer} &=& \left(\sigma_{\textrm{layer},x}(\xi),
  \,\,\sigma_{y}(\theta), \,\,3.46\,\, \textrm{mm}\right).
\end{eqnarray}
The error in a cluster's position mainly depends on the dip angle $\theta$ and the crossing angle $\xi$
which are shown in Fig.~\ref{figure:hit_cluster_position_error}.
The effect of electron diffusion is expected to be enhanced at large $\theta$
because the effect on neighboring pads increases the width of the pulse shape.
The same effect is expected for $\xi$ as it approaches the angle
$\left|\xi\right| = 45^\circ$ and $\left|\xi\right| = 135^\circ$
because hit clusters are formed in the plane of $\xi=0$ for row-clusters and $\xi=90^\circ$ for layer-clusters.
On the other hand, saturation of $\sigma_{\textrm{row},z}$ at $\xi>65^\circ$ and $\sigma_{\textrm{layer},x}$
at $\xi<10^\circ$ is seen which is an effect of the pad response function~\citep{estee2019162509}.
The constant components $\sigma_{\textrm{layer},z}$ and $\sigma_{\textrm{row},x}$ come from the
standard deviation of the uniform distribution where the width of the distribution is the width of the pad.

The energy loss per unit length, $dE/dx$ (ADC/mm), can be calculated for each hit cluster
that is used for the particle identification analysis discussed in section~\ref{pid}.
The energy loss $dE$ is defined as the hit cluster charge $Q$
and the length $dx$ is calculated from the path length of the track
in the cluster box range as shown in Fig.~\ref{figure:clusterization}.

%%%%%%%%%%%%%%%%%%%%%%%%%%%%%%%%%%%%%%%%%%%%%%%%%%%%%%%%%%%%%%%%%%%%%%%%%%%%%%%%%%%%%%%%%%%%%%%%%%
%%%%%%%%%%%%%%%%%%%%%%%%%%%%%%%%%%%%%%%%%%%%%%%%%%%%%%%%%%%%%%%%%%%%%%%%%%%%%%%%%%%%%%%%%%%%%%%%%%
\section{Momentum reconstruction}
\begin{figure}[ht]
  \centering\includegraphics[width=6cm]{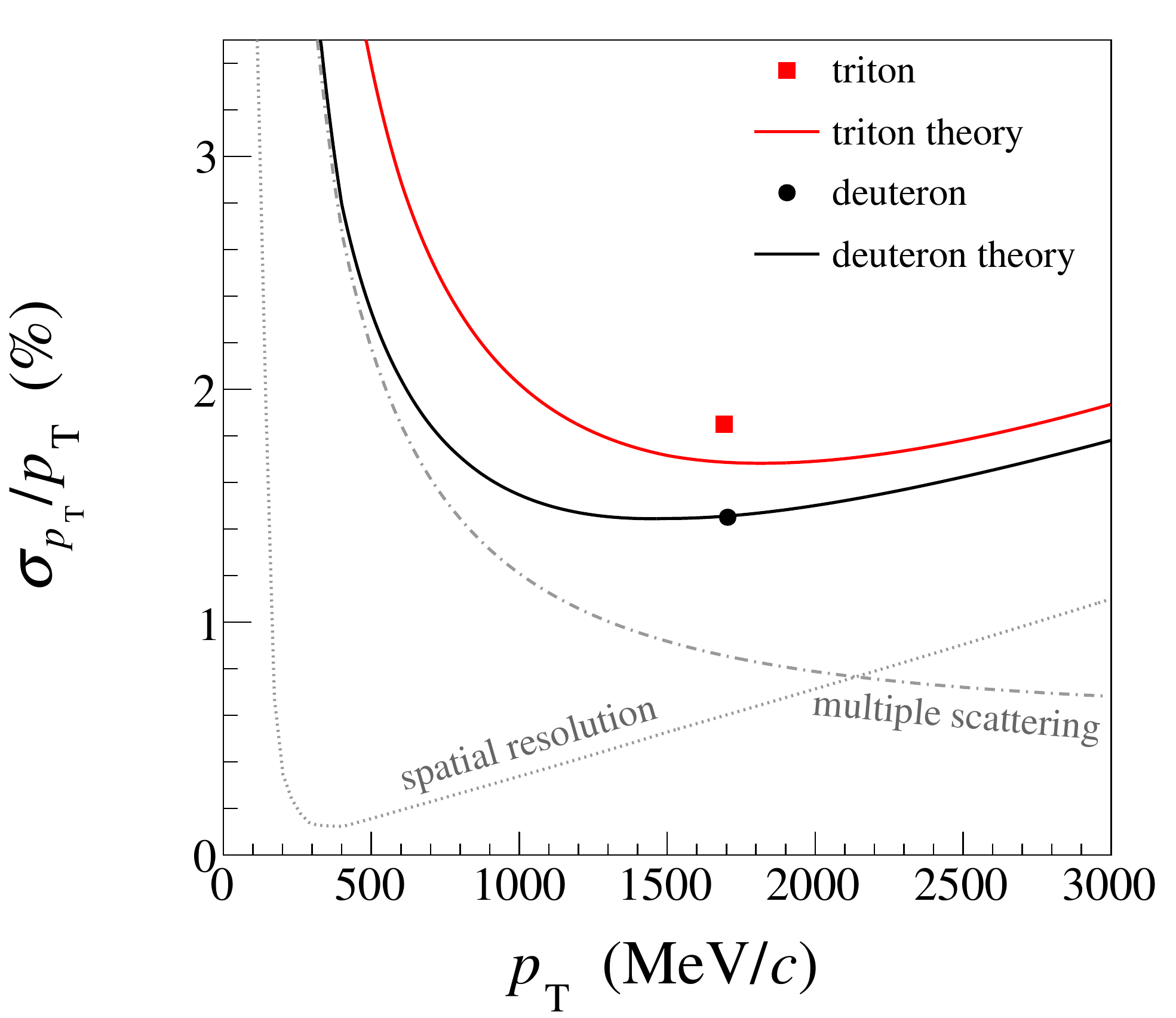}
  \caption{Transverse momentum resolution of deuteron and triton.
  The theoretical lines as a function of momentum are compared with the data.
  The two factors of momentum resolution (spatial resolution and multiple scattering)
  are indicated for the deuteron.
  }
\label{figure:momentum_resolution}
\end{figure}

The momentum reconstruction is performed on the hit clusters using GENFIT~\citep{rauch2015}. GENFIT is able to perform fits on the hit clusters in 3-dimensions with a Kalman fitter algorithm~\citep{FRUHWIRTH1987444}, stepping through the various materials while calculating the energy loss and next point along the way leading to momentum reconstruction.
The hit cluster position along with its error, gas properties and magnetic field map availiable in ref.~\citep{fieldmap}
are provided as input.
The accuracy in determining the particle momentum from the reconstructed tracks is affected by both the $E \times B$ effect due to the magnetic field inhomogeneity and the space charge caused by unreacted beam particles traversing the active gas volume of the TPC~\citep{tsang2019space}. These effects are best evaluated with Monte Carlo simulations which are currently underway and will be the subject of a future publication. Nonetheless, we can  evaluate the transverse momentum resolution using a cocktail particle beam of deuterons and tritons at momentum around 1700 MeV/$c$ taken during the experiment.
The resolution is 1.5 \% for deuteron and 2 \% for tritons as shown by the symbols in Fig.~\ref{figure:momentum_resolution}.
The measured resolution is very closed to the predicted values as shown by the solid curves.

\begin{figure}[ht]
  \centering\includegraphics[width=6cm]{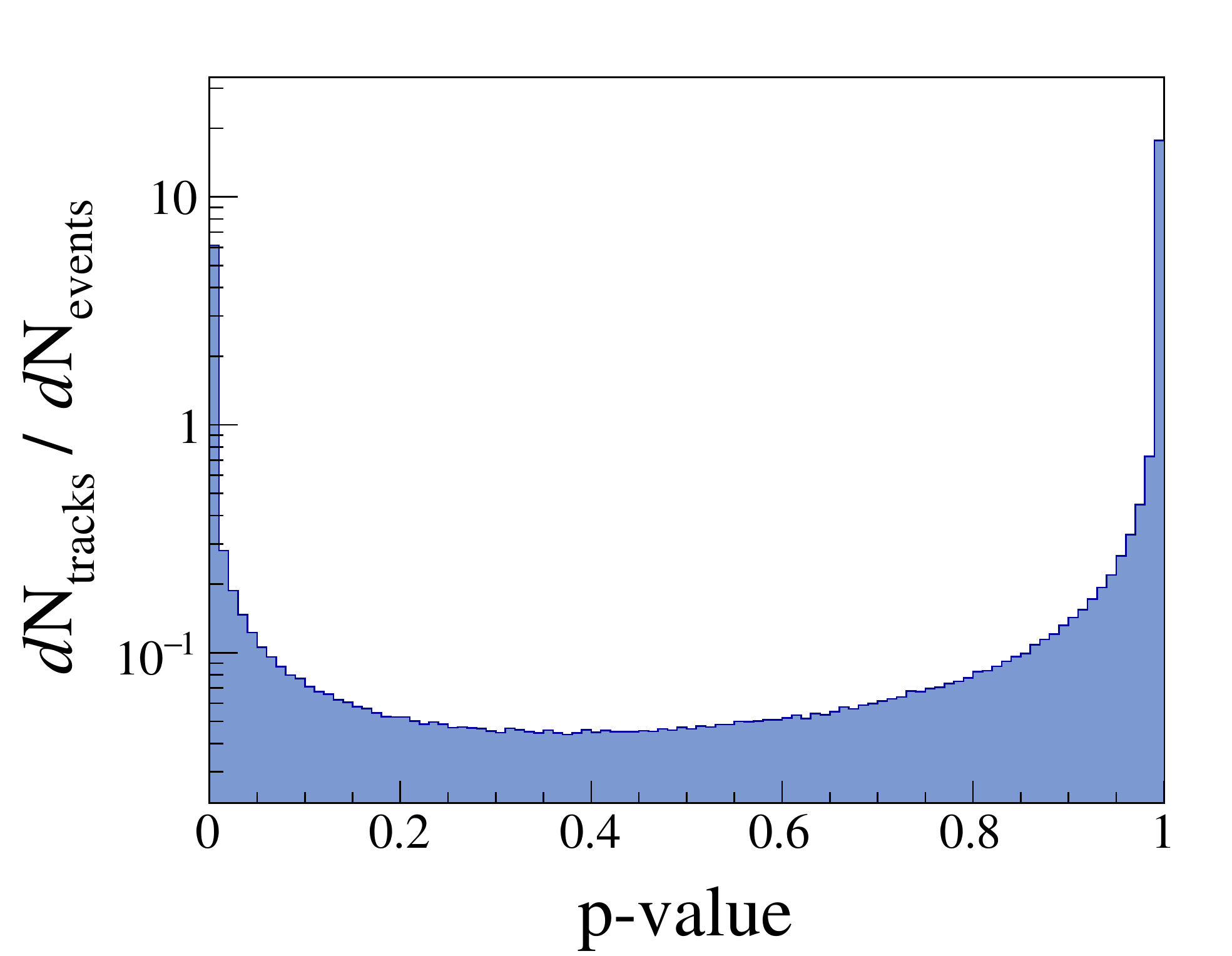}
  \caption{The p-value distribution of the track fit.}
  \label{figure:pval}
\end{figure}

\begin{figure}[ht]
  \centering\includegraphics[width=6cm]{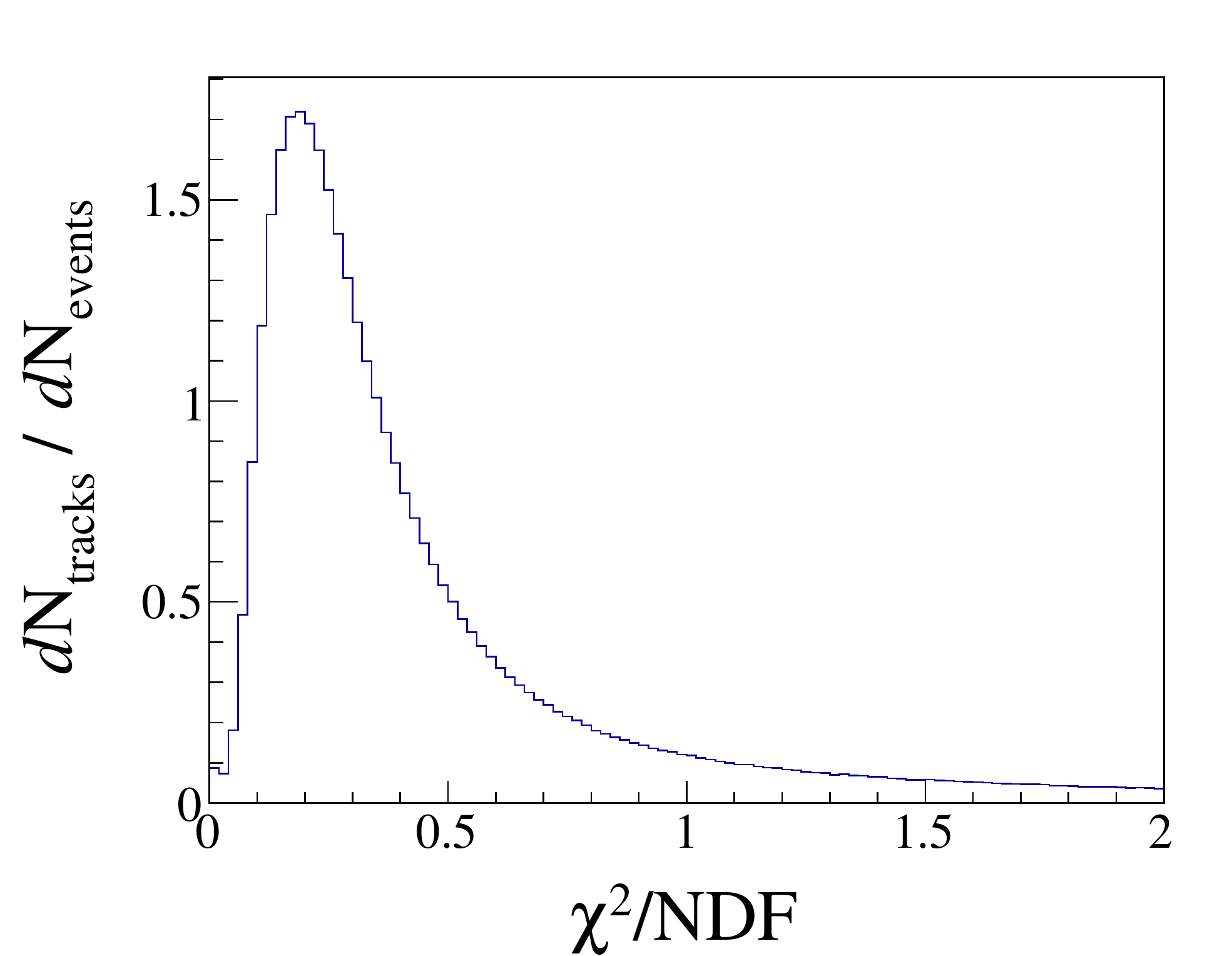}
  \caption{The $\chi^2/\textrm{NDF}$ distribution of the track fit.}
  \label{figure:chi2ndf}
\end{figure}

\begin{figure}[hb]
  \centering\includegraphics[width=6cm]{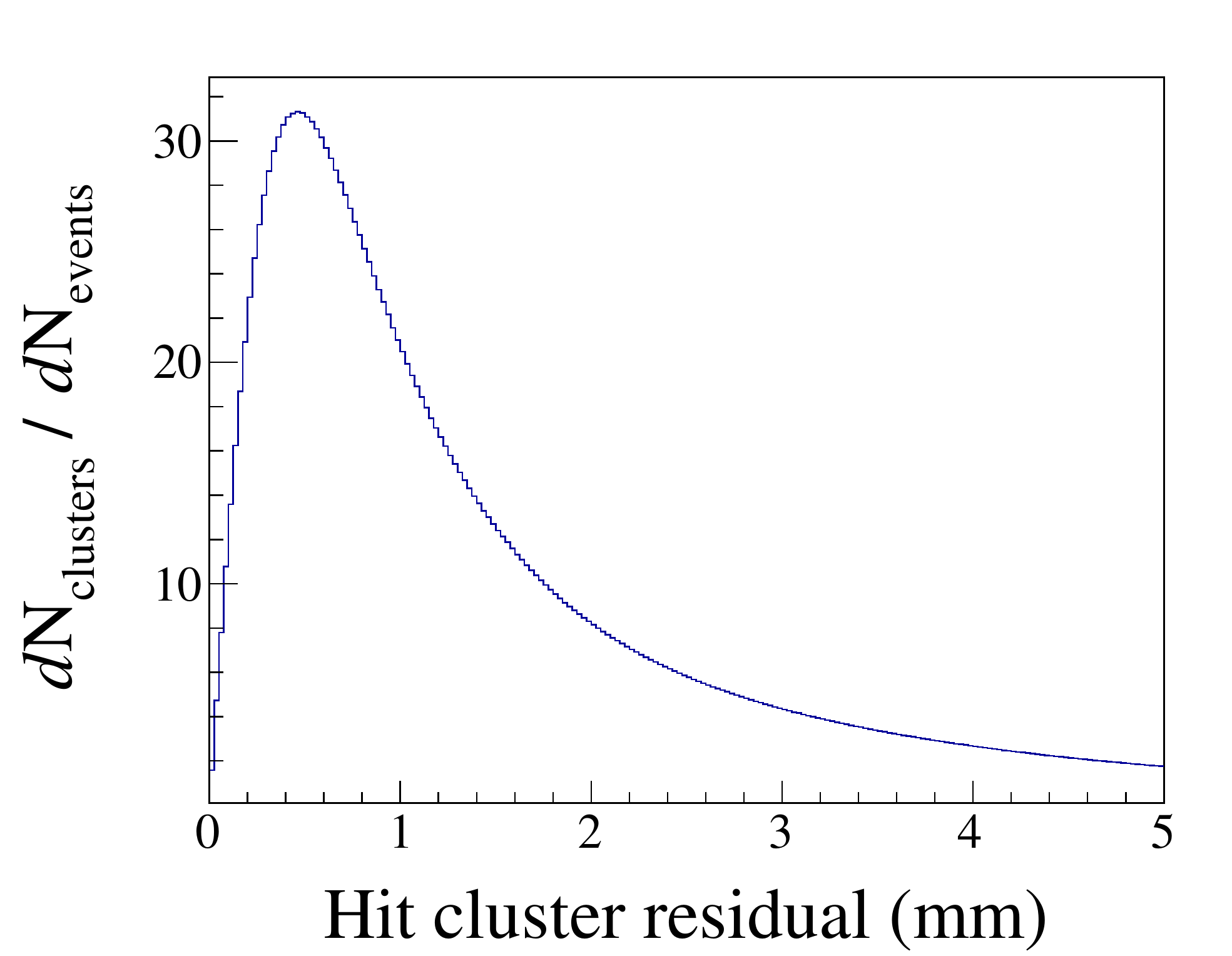}
  \caption{The hit cluster residual of tracks after they are fit using GENFIT.}
\label{figure:hitcluster_residual}
\end{figure}

The track fit is verified with the p-value, the $\chi^2$ and the Number of Degrees of Freedom (NDF).
The p-value is the probability transform of the track fit defined by the cumulative distribution function of the $\chi^2$:
\begin{equation} \label{eq:pvalue}
  \textrm{p-value} = \int^{\infty}_{\chi^2}
  \frac{x^{m\!-\!1}\textrm{e}^{\!-x\!/\!2}}{\Gamma(m)2^{m}}dx,
\end{equation}
where $m=\textrm{NDF}/2$.
The distribution of p-value and $\chi^2/\textrm{NDF}$ are shown in Fig.~\ref{figure:pval} and \ref{figure:chi2ndf}, respectively.
The resulting hit cluster residual is shown in Fig.~\ref{figure:hitcluster_residual}.

\section{Vertex reconstruction}
Determining position information for the primary vertices is essential for selecting on-target events, selecting primary tracks, and improving momentum resolution. The
%{C\nolinebreak\hspace{-.05em}\raisebox{.4ex}{\tiny\bf +}\nolinebreak\hspace{-.10em}\raisebox{.4ex}{\tiny\bf +}\ }
C++
package \linebreak RAVE~\citep{waltenberger2011} is used for vertex reconstruction in the S$\pi$RIT-TPC. RAVE is the optimized vertex reconstruction tool for high multiplicity events developed by the CMS community~\citep{cms2008}. We used the Adaptive Vertex Fitter (AVF)~\citep{waltenberger2007} among other fitters for the reconstruction of a vertex. The AVF is an iterative weighted Kalman filter, which will assign a weight to each track, ignoring outlier tracks. The user parameters for the implementation of RAVE are adjusted as explained in Ref.~\citep{waltenberger2011}.

\begin{figure}[!hbt]
  \centering\includegraphics[width=8cm]{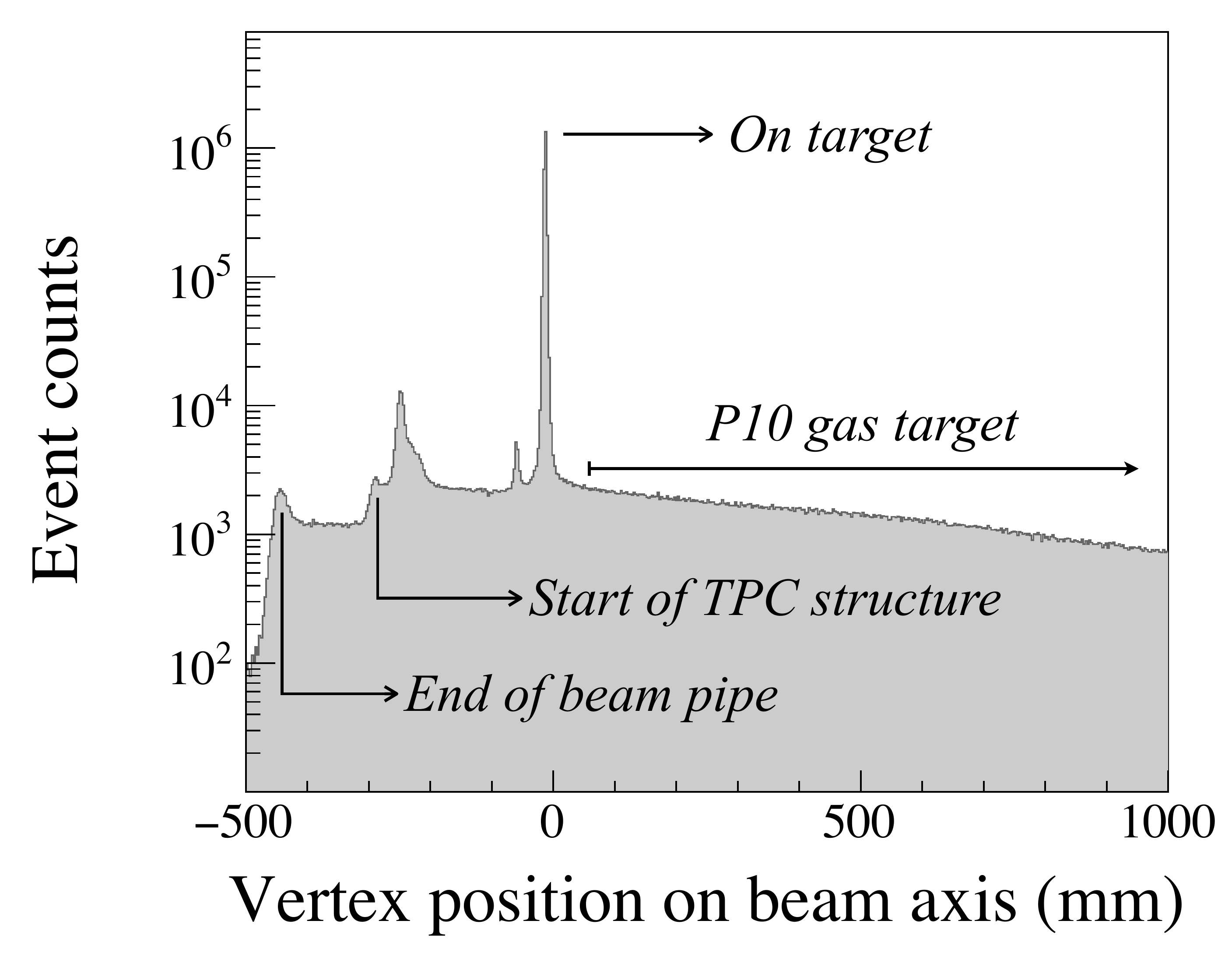}
  \caption{Vertex distribution along the beam axis. The beam pipe window, TPC structure, and on-target collision events are clearly evident. The decreasing trend of P10 gas target events is due to the trigger condition provided by the Kyoto Multiplicity Array.}
  \label{figure:vertex_position}
\end{figure}

The position distribution of reconstructed vertices along the beam axis is shown in Fig.~\ref{figure:vertex_position}. Peaks corresponding to the position of the beam pipe window, the entrance of the TPC structure, and the target are clearly visible, followed by active-target type events in the P10 gas. The occurrence of active-target events decreases along the beam axis, as a result of the trigger condition set by the Kyoto Multiplicity Array~\citep{kaneko2005} which consists of a total of 60 scintillators, placed on both sides of the S$\pi$RIT-TPC.

\begin{figure}[!hbt]
  \centering\includegraphics[width=6cm]{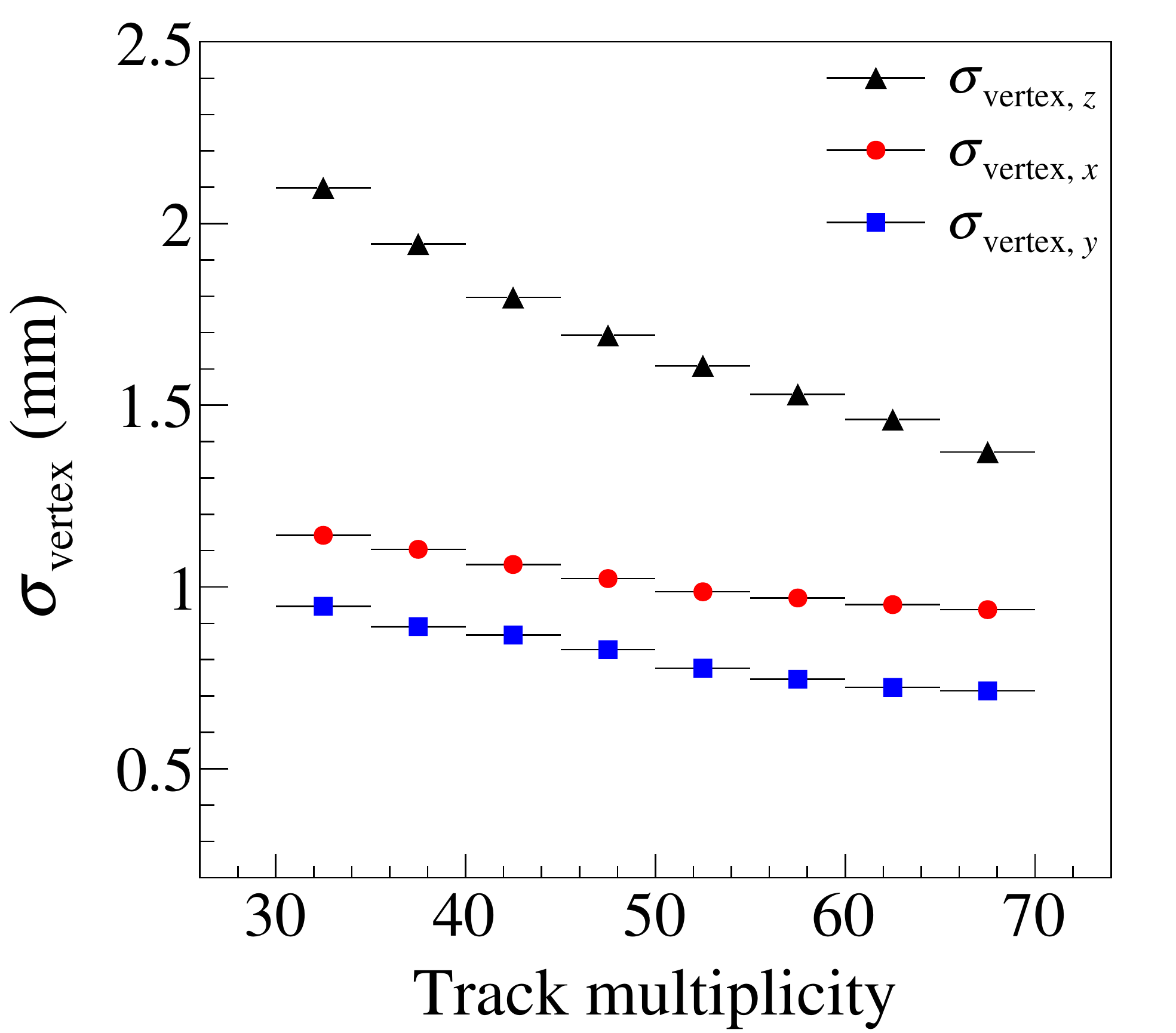}
  \caption{Vertex resolutions as a function of the track multiplicity.
  The overall resolution improves as the track multiplicity increases.}
  \label{figure:vertex_resolution}
\end{figure}

The target vertex resolution is plotted as a function of the track multiplicity in Fig.~\ref{figure:vertex_resolution}. Note that the target thickness in the $z$-axis is 0.8 mm. The reconstructed vertex position from RAVE for on-target events was used to obtain the $z$ component of the position resolution $\sigma_{\textrm{vertex},z}$, while the transversal components of the position resolution $\sigma_{\textrm{vertex},x}$ and $\sigma_{\textrm{vertex},y}$ was evaluated by comparing with data from beam line detectors. The overall position resolution of the primary vertex improves with track multiplicity.

\begin{figure}[ht]
  \centering\includegraphics[width=6cm]{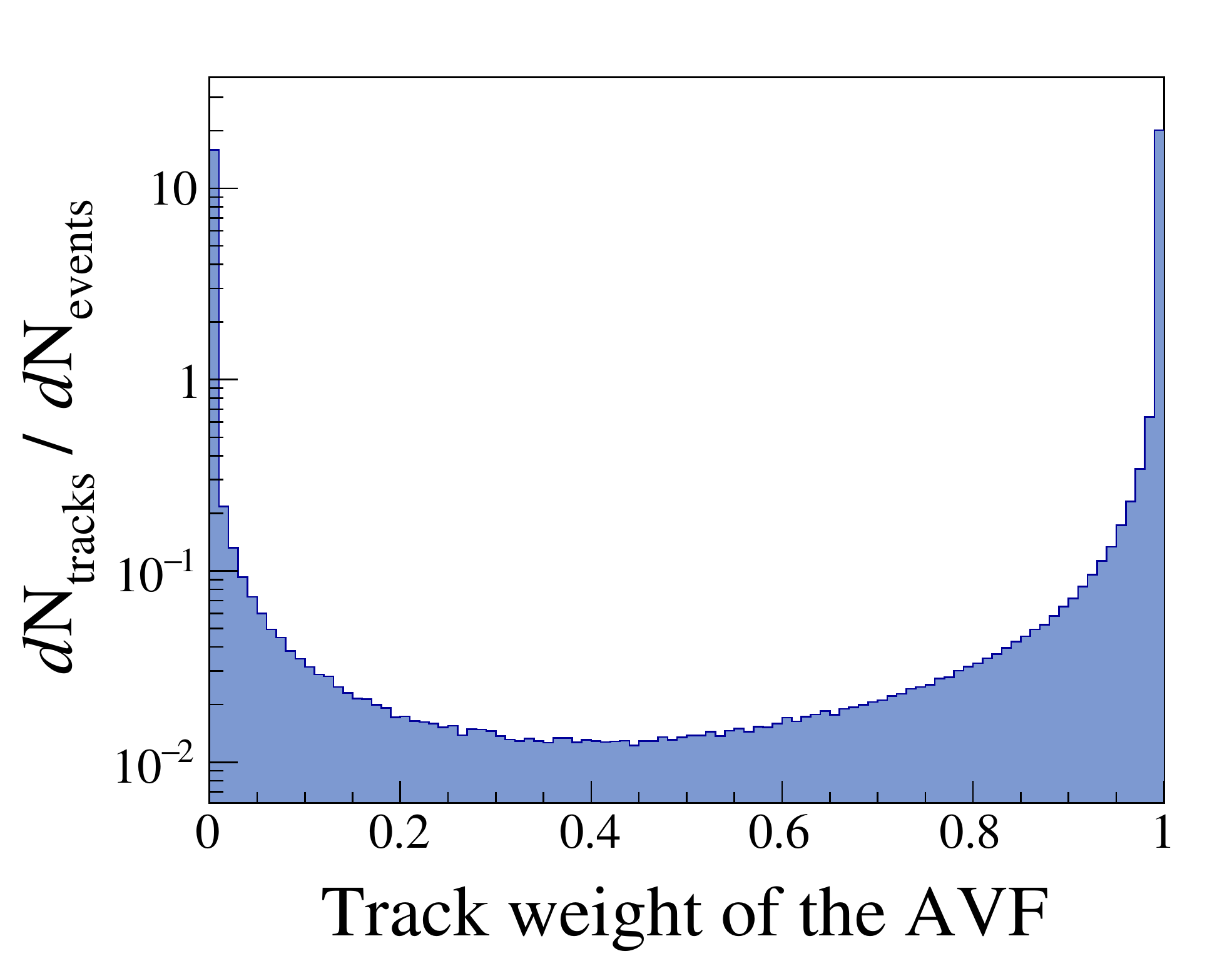}
  \caption{The distribution of the track weight from the AVF.}
  \label{figure:w}
\end{figure}

As a probability projection of tracks belonging to the reconstructed vertex, the track weight of the AVF
is shown in Fig.~\ref{figure:w}.
The dependency on the track multiplicity is negligible.

%%%%%%%%%%%%%%%%%%%%%%%%%%%%%%%%%%%%%%%%%%%%%%%%%%%%%%%%%%%%%%%%%%%%%%%%%%%%%%%%%%%%%%%%%%%%%%%%%%
%%%%%%%%%%%%%%%%%%%%%%%%%%%%%%%%%%%%%%%%%%%%%%%%%%%%%%%%%%%%%%%%%%%%%%%%%%%%%%%%%%%%%%%%%%%%%%%%%%
\section{Particle Identification} \label{pid}
To identify particle species, the truncated mean of energy loss per unit length, $\left<dE/dx\right>$, and
the magnetic rigidity, $p/Z$ (momentum divided by charge), is used.
The $dE/dx$ distribution of charged particles in the gas can be described by the
Landau function which is also known as the straggling function~\citep{bichsel2006}.
The most reliable parameter to represent the $dE/dx$ distribution of the tracks is the most-probable value.
However, the number of hit clusters is usually not enough to
reconstruct the complete $dE/dx$ distribution function.
Moreover, rare single-collisions sometimes leave with high energy loss,
and it distorts the straggling function.
Therefore, the truncated mean of $dE/dx$ is preferred to the most-probable value in the TPC analysis.
For S$\pi$RIT-TPC, the mean energy loss per unit length $\left<dE/dx\right>$ is calculated by
truncating the highest $30\,\%$ of the $dE/dx$ values of hit clusters.
The disadvantage of this method is that the total $\left<dE/dx\right>$ spectrum does not follow the
empirical Bethe-Bloch formula when compared to the particles with different charges,
because the width of the straggling function is
significantly modified depending on the charge number $Z$.

\begin{figure*}[ht]
  \centering\includegraphics[width=10cm]{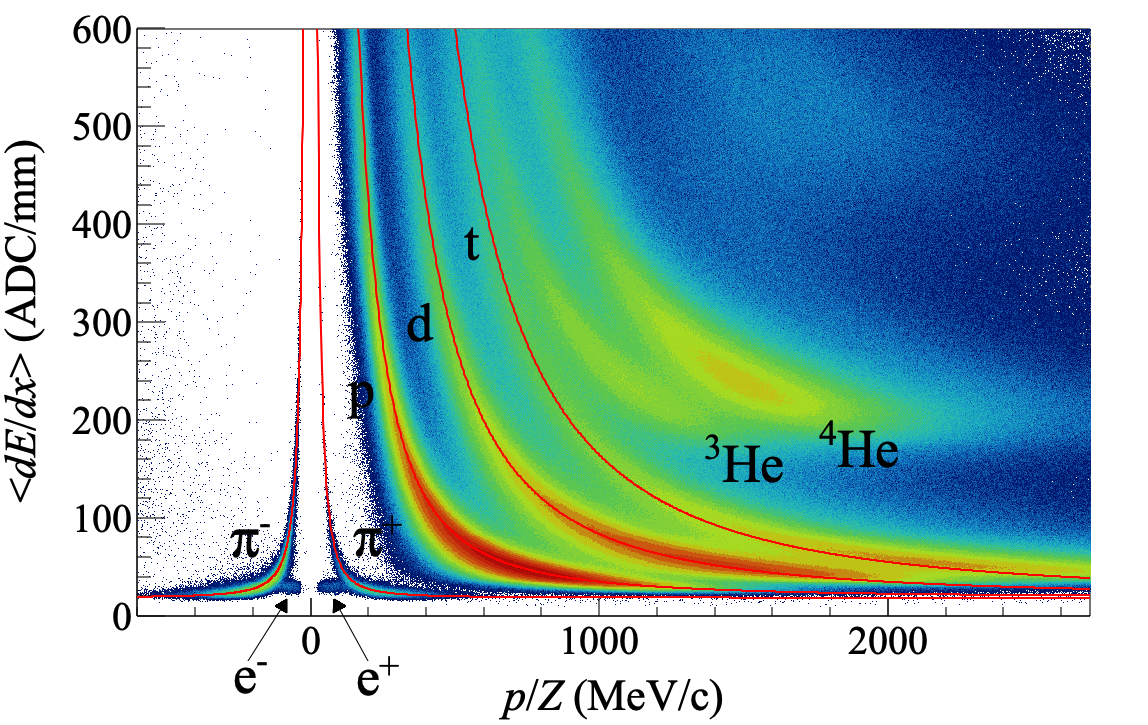}
  \caption{$\left<dE/dx\right>$ spectrum vs. magnetic rigidity $p/Z$.
  The distribution of particles is compared with the empirical Bethe-Bloch formula shown in red lines.}
  \label{figure:pid_plot}
\end{figure*}

In correlation plot between $\left<dE/dx\right>$ and magnetic rigidity,
pions, hydrogen isotopes,
and helium isotopes can be identified.
The clouds of electrons and positrons are also observed below the $\pi^{\pm}$ distributions.
The red curves in Fig.~\ref{figure:pid_plot} is guided with the empirical Bethe-Bloch formula.
The proton spectrum is used to determine the fit parameters in the Bethe-Bloch formula
which are then extrapolated to other $Z=1$ particles.

%%%%%%%%%%%%%%%%%%%%%%%%%%%%%%%%%%%%%%%%%%%%%%%%%%%%%%%%%%%%%%%%%%%%%%%%%%%%%%%%%%%%%%%%%%%%%%%%%%
%%%%%%%%%%%%%%%%%%%%%%%%%%%%%%%%%%%%%%%%%%%%%%%%%%%%%%%%%%%%%%%%%%%%%%%%%%%%%%%%%%%%%%%%%%%%%%%%%%
\section{Summary}
The framework software S$\pi$RITROOT has been developed
to reconstruct and to analyze heavy ion collision data taken with the S$\pi$RIT TPC at RIBF, RIKEN, Japan.
In this software, multi-pulse fitting, fast helix fitting, and fast POCA calculation are incorporated.
The momentum reconstruction package GENFIT and the vertex reconstruction package RAVE
have been successfully adapted to the box-type detector system.
The position distribution of the reconstructed vertices reflects the structure of the experimental
setup and the on-target events.
In the correlation plot of energy loss and magnetic rigidity,
pions, hydrogen isotopes,
and helium isotopes are identified.
The estimation of the reconstruction efficiencies
of the particles using a Monte-Carlo simulation
will be an important task for the future.

\section{Acknowledgments}
This work was supported by
the U.S. Department of Energy under Grant Nos. DE-SC0014530, DE-NA0002923,
US National Science Foundation Grant No. PHY-1565546,
the Japanese MEXT KAKENHI (Grant-in-Aid for Scientific Research on Innovative Areas) grant No. 24105004,
and the National Research Foundation of Korea (NRF) under grant Nos.\linebreak 2016K1A3A7A09005578 and 2018R1A5A1025563.
The authors are indebted to RIBF technical staffs for excellent beams.
The computing resources for analyzing data were provided by the HOKUSAI-GreatWave system at RIKEN.

\biboptions{square,sort&compress,comma,numbers}
\bibliographystyle{model1-num-names.bst}
\bibliography{tracking_software.bib}

\end{document}